\documentclass[12pt, landscape, twocolumn]{article}

\usepackage{amsmath, amssymb, amsthm}
\usepackage{graphicx}
\usepackage{hyperref}
\usepackage{geometry}
\usepackage{bbm}
\usepackage{cancel}
\usepackage{lineno}
\usepackage{xurl}
\usepackage[strings]{underscore}  %

\usepackage{tikz}
\usetikzlibrary{arrows.meta, positioning, shapes.geometric}

\usepackage{fancyhdr}
\newtheorem{theorem}{Theorem}[section]
\newtheorem{definition}[theorem]{Definition}

\newcommand{\bR}{\mathbb{R}}
\newcommand{\bZ}{\mathbb{Z}}
\newcommand{\bC}{\mathbb{C}}
\newcommand{\bN}{\mathbb{N}}

\newcommand{\bE}{\mathbb{E}}

\newcommand{\cL}{\mathcal{L}}

\newcommand{\cO}{\mathcal{O}}
\newcommand{\cN}{\mathcal{N}}

\newcommand{\cD}{\mathcal{D}}
\newcommand{\cA}{\mathcal{A}}
\newcommand{\cS}{\mathcal{S}}
\newcommand{\cV}{\mathcal{V}}
\newcommand{\cB}{\mathcal{B}}

\newcommand{\cZ}{\mathcal{Z}}
\newcommand{\teq}[1]{\stackrel{\text{#1}}{=}}
\newcommand{\din}{d}
\newcommand{\bk}[1]{\langle #1 \rangle}
\newcommand{\bbk}[1]{\big\langle #1 \big\rangle}

\newcommand{\maps}{\text{Maps}}

\definecolor{ForestGreen}{cmyk}{0.91,0,0.88,0.12} %

\newcommand{\Q}[1]{
    \begin{center}
        \vspace{.25cm}
    \begin{minipage}{.45 \textwidth}
        \textbf{Question:} #1
        \vspace{.25cm}
    \end{minipage}
    \end{center}
    }

\newcommand{\A}[1]{
    \begin{center}
        \vspace{.25cm}
    \begin{minipage}{.45 \textwidth}
        \textbf{Answer:} #1
        \vspace{.25cm}
    \end{minipage}
    \end{center}
    }

\newcommand{\C}[1]{
    \begin{center}
        \vspace{.25cm}
    \begin{minipage}{.45 \textwidth}
        #1
    \end{minipage}
    \vspace{.25cm}
    \end{center}
    }

\newcommand{\CT}[2]{
    \begin{center}
        \vspace{.25cm}
    \begin{minipage}{.45 \textwidth}
        \textbf{#1:} #2
        \vspace{.25cm}
    \end{minipage}
    \end{center}
    }

\renewcommand{\O}[1]{
    \begin{center}
        \vspace{.25cm}
    \begin{minipage}{.45 \textwidth}
        \textbf{Observation:} #1
    \end{minipage}
    \vspace{.25cm}
    \end{center}
    }

\title{ \bf \Huge Pre-Strings Lectures on Artificial Intelligence}
\author{\LARGE Jim Halverson\vspace{.75cm}\\ \emph{Department of Physics, Northeastern University, Boston, MA 02115, USA}\vspace{.3cm}\\ \emph{The NSF Institute for Artificial Intelligence} \\ \emph{and Fundamental Interactions}\vspace{.3cm} \\ \emph{Simons Collaboration on the Physics of Learning} \\ \emph{and Neural Computation} \vspace{.3cm}\\ \emph{Physical Superintelligence, PBC} \vspace{.3cm}\\ \texttt{jhh@neu.edu}}
\date{}

\begin{document}

\onecolumn

\maketitle
\thispagestyle{empty}

\begin{center}
        \begin{minipage}{0.75\textwidth}
        \begin{abstract}
            These notes are based on six lectures given over three days at the Pre-Strings 2026 school in Shanghai. Day~1 develops neural network essentials, organized around the expressivity, statistics, and dynamics of neural networks, presented with a field-theoretic lens. Day~2 develops a neural network approach to field theory (NN-FT), in which a field theory is defined by a network architecture and a density on its parameters, and surveys recent results. Examples include a universality theorem, a neural network realization of Liouville theory, famous string amplitudes, topological sectors and the Kosterlitz-Thouless transition, Ward identities and anomalies, and a new  derivation of the critical dimension of the bosonic string. Day~3 turns the lens around and covers applied AI for string theory: agentic workflows that are changing how the other techniques are implemented, physics-informed neural networks and Calabi-Yau metrics, reinforcement learning and search in the string landscape and in knot theory, and interpretable supervised learning with an eye towards conjecture generation.
        \end{abstract}
        \end{minipage}

\end{center}

\clearpage
\twocolumn
\tableofcontents
\clearpage

\section{Introduction}
\label{sec:lecture_intro}

I opened my TASI 2024 lectures \cite{halverson2024tasilecturesphysicsmachine} with a provocation: computer science is still in its infancy. The argument was historical. From Lovelace and Babbage and Turing, through code-breaking in the 40's, personal computers in the 70's and 80's, the internet in the 90's, supercomputers-in-your-pocket in the 2000's, and some version of artificial intelligence in the 2010's \cite{isaacson2014innovators}, the field has compounded for decades and shows no signs of slowing. A future historian, I wagered, will measure progress in CS in centuries, not decades, and from that vantage point we are at the very beginning.

It is July 2026, and the progress of the last two years has made the provocation look conservative. Let me tell you what changed.

First, language models learned to \emph{reason}. The LLMs of June 2024 were already impressive, but they answered in a single pass. The new generation of \textbf{reasoning models} \cite{openai2024o1, deepseek2025r1} is instead trained with reinforcement learning on long chains of thought: the model writes out its reasoning, explores, backtracks, checks itself, and only then answers, with performance that improves the longer it is allowed to think. This is a dynamics story, RL on the space of reasoning trajectories, and it led to progress on benchmarks that had seemed comfortably out of reach. By July 2025, a version of Gemini with Deep Think and an experimental OpenAI reasoning model both achieved gold-medal level scores at the International Mathematical Olympiad \cite{deepmind2025imo, openai2025imo}, solving five of six problems under contest conditions. By early 2026, language models had produced the first solutions of Erd\H{o}s problems \cite{erdosproblems} that were not previously in the literature, with proofs formalized in Lean and checked by expert mathematicians; and by May 2026 an OpenAI reasoning model \cite{openai2026unitdistance} had generated a counterexample disproving Erd\H{o}s's 1946 unit distance conjecture, digested and verified in a short companion paper by nine mathematicians \cite{alon2026remarks}.

Second, models learned to \emph{act}. A reasoning model wrapped in a loop (think, call a tool, observe the result, think again \cite{yao2023react}) is an \textbf{agent}, and agents now write and execute code, search the literature, run computations, criticize their own outputs, and orchestrate teams of subagents over hours- or days-long tasks (e.g., \cite{claudecode}). Evolutionary agentic systems have already made discoveries: FunSearch \cite{romera2024funsearch} found new constructions in extremal combinatorics, and AlphaEvolve \cite{novikov2025alphaevolve} discovered a novel procedure for a type of matrix multiplication, the first improvement in many years. Closer to home, agents are increasingly woven into the daily practice of theoretical physics, including for some of the results presented in Day 2 of these lectures, where AI-assisted research workflows accelerated parts of the analysis (with the humans, as always, responsible for correctness). We will return to agents at length in Day~3, where I will argue they have lowered the cost of every ML-for-physics technique in these lectures.

\bigskip
These lectures were given at the Pre-Strings 2026 school \cite{prestrings2026} at SIMIS, preceding Strings 2026, to an audience of Ph.D. students and postdocs in string theory. I had three days, which I used as follows:
\begin{itemize}
\item \textbf{Day 1: Neural Network Essentials.} A theorist's introduction to neural networks, organized around three pillars: expressivity (including transformers and reasoning), statistics, and dynamics. The core material was developed for \cite{halverson2024tasilecturesphysicsmachine}.
\item \textbf{Day 2: Neural Network Field Theory.} A new approach to field theory motivated by ideas from ML theory. After covering the essentials, the second half covers results from the last year: e.g., universality, Liouville theory, the bosonic string, topological defects and the Kosterlitz-Thouless transition, and anomalies.
\item \textbf{Day 3: Applied AI for Strings.} The ML-for-physics direction, accelerated by agents. Topics include physics-informed neural networks and Calabi-Yau metrics, reinforcement learning and search in physics landscapes, and interpretable supervised learning aimed at conjecture generation.
\end{itemize}
The days build on one another: Day 1's essentials are the working language of Day 2, and Day 3 turns from understanding neural networks to using them on string theory problems. Throughout I take a decidedly field-theoretic lens, to suit the audience and because I think it is a useful way to understand neural networks.

\section{Day 1: Neural Network Essentials}
\label{sec:day1}

\subsection{The Setup}
\label{sec:setup}

If we want to understand machine learning, we must at minimum understand neural networks. A neural network is a function
\begin{equation}
\phi_\theta:\bR^d \to \bR
\end{equation}
depending on parameters $\theta$. Scalar outputs suffice for the essential points we wish to make, but indices can be added according to the shape of the data, to encode images, sequences, etc. We fix some vocabulary:
\begin{align}
    \text{Input:} & \quad x \in \bR^d \\
    \text{Output:} & \quad \phi_\theta(x) \in \bR \\
    \text{Network:} & \quad \phi_\theta \in \maps(\bR^d, \bR) \\
    \text{Data:} & \quad \cD,
\end{align}
where the data $\cD$ is problem-dependent but we imagine as a subset of $\bR^d$, possibly with labels $y \in \bR$ attached.

With only this much structure we can already pose the central question of today's lectures:
\Q{What does a NN predict?}
At fixed $\theta$ the answer is trivially $\phi_\theta(x)$. But this answer is na\"ive, for reasons of dynamics and of statistics.

Dynamics first. Learning means updating parameters, so the objects of interest are really \textbf{trajectories} in
\begin{align}
    \text{Parameter Space:} & \quad \theta(t) \in \bR^{| \theta |} \\
    \text{Output Space:} & \quad \phi_{\theta(t)}(x) \in \bR \\
    \text{Function Space:} & \quad \phi_{\theta(t)}\in \maps(\bR^d, \bR),
\end{align}
generated by a learning algorithm acting on a problem. In supervised learning, for instance, the data is
\begin{equation}
\cD = \{(x_\alpha, y_\alpha) \in \bR^d \times \bR\}_{\alpha=1}^{|\cD|},
\end{equation}
one builds a loss functional
\begin{equation}
\cL[\phi_\theta] = \sum_{\alpha=1}^{|\cD|} \ell(\phi_\theta(x_\alpha), y_\alpha),
\end{equation}
with $\ell$ a per-sample loss such as $\ell_{\text{MSE}} = (\phi_\theta(x_\alpha)-y_\alpha)^2$, and one descends the gradient
\begin{equation}
    \frac{d\theta_i}{dt} = -\nabla_{\theta_i} \cL[\phi_\theta],
\end{equation}
or runs a cousin of gradient descent such as SGD \cite{sgd,perceptron}, Adam \cite{kingma2017adammethodstochasticoptimization}, or a physics-inspired alternative like Energy Conserving Descent \cite{de2022born,de2023improving}. Unless stated otherwise, $t$ always denotes training time.

Now statistics. When your computer initializes a network, the parameters are sampled,
\begin{equation}
    \theta \sim P(\theta),
\end{equation}
and different draws of parameters give different functions. No single draw is preferred over any other, so the pointwise prediction $\phi_\theta(x)$ of one randomly initialized network simply cannot be fundamental. What is fundamental is the ensemble: the mean prediction, the second moment,
\begin{align}
    \bE[\phi_\theta(x)] & = \int d\theta \,P(\theta) \, \phi_\theta(x) \\
    \bE[\phi_\theta(x) \phi_{\theta}(y)] & = \int d\theta \, P(\theta)  \, \phi_\theta(x) \phi_{\theta}(y),
\end{align}
and the higher moments, where the expectation runs over initializations. We are physicists, so we write $\bE[\cdot]=\bk{\cdot}$ and recognize old friends:
\begin{align}
    G^{(1)}(x) & = \bk{\phi_\theta(x)} \\
    G^{(2)}(x, y) & = \bk{\phi_\theta(x) \phi_\theta(y)}.
\end{align}
The mean prediction and second moment of the network ensemble are the one-point and two-point correlation functions. Already we see that field theoretic language appears in ML.

Combining the dynamics and statistics: each initialization seeds a trajectory, so we have an ensemble of trajectories, described by a time-dependent density
\begin{equation}
    \theta(t) \sim P_t(\theta)
\end{equation}
and time-dependent correlators
\begin{align}
    G_t^{(1)}(x) & = \bk{\phi_\theta(x)}_t \\
    G_t^{(2)}(x, y) & = \bk{\phi_\theta(x) \phi_\theta(y)}_t,
\end{align}
evaluated in the time-evolving density on parameters.
If training is doing its job we care especially about late times, e.g.
\C{$G_\infty^{(1)}(x)$ = mean prediction of $\infty$-many NNs as $t\to\infty$,}
and remarkably we will see that in a particular supervised setting this quantity is exactly solvable.

One more pillar is needed. Consider the architecture
\begin{equation}
\phi_\theta(x) = \theta \cdot x.
\end{equation}
This is a linear model, learning with it is linear regression, and it will fail on any interesting non-linear problem. The model is not \emph{expressive} enough. We need architectures rich enough to approximate essentially anything, which in particular forces non-linearity.

\medskip
We therefore study neural networks through three \textbf{pillars}, each with its own question: 
\begin{itemize}
    \item {\bf Expressivity.} How powerful is the NN?
    \item {\bf Statistics.} What is the NN ensemble?
    \item {\bf Dynamics.} How does it evolve?
\end{itemize}
A physicist's lens means physicist's tools (field theory for the ensembles, landscape dynamics for the loss, symmetries in various contexts) and a physicist's standard of rigor: mechanisms and toy models over theorems, intuition with control. This befits a field where experiment is far ahead of theory and what is needed first is $O(1)$ understanding of the empirical phenomena.

Henceforth we usually drop the subscript and write $\phi(x)$ for $\phi_\theta(x)$; the parameter-dependence is always there.

\subsection{Expressivity of Neural Networks}
\label{sec:expressivity}

Neural networks are big functions built by composing many small ones, with the composition pattern fixed by the architecture. Composition buys flexibility, so we ask
\Q{How powerful is a NN?}
This is fundamentally a math question  about approximating functions, and the classic answers are approximation theorems.

\subsubsection{Universal Approximation Theorem}

Perhaps the most famous of these results is the Universal Approximation Theorem (UAT): one hidden layer suffices to approximate any continuous function on a compact set to any desired accuracy. Cybenko's original version reads:
\begin{theorem}[Cybenko \cite{cybenko1989approximation}]
Let $f: \bR^d \to \bR$ be a continuous function on a compact set $K\subset \bR^d$. Then for any $\epsilon>0$ there exists a neural network with a single hidden layer of the form
\begin{equation}
\phi(x) = \sum_{i=1}^N  w_i^{(1)} \sigma\left(\sum_{j=1}^d w_{ij}^{(0)}  x_j + b^{(0)}_i\right)+b^{(1)}, \label{eqn:perceptron}
\end{equation}
$\theta = \{w_{ij}^{(0)}, w_i^{(1)}, b^{(0)}_i, b^{(1)}\}$,
where $\sigma:\bR \to \bR$ is a non-polynomial non-linear activation function, such that
\begin{equation}
\sup_{x\in K} |f(x) - \phi(x)| < \epsilon.
\end{equation}
\end{theorem}
\noindent The architecture \eqref{eqn:perceptron} goes by many names: perceptron, fully-connected network, {\bf feedforward network}. The integer $N$ is the {\bf width}. Stacking affine maps and non-linearities $L$ times adds a {\bf depth} dimension, giving the multi-layer perceptron (MLP), or deep feedforward network.

Powerful as it is, the UAT has two blind spots worth internalizing. It is silent on \emph{how many} neurons a given accuracy requires, and it is silent on \emph{how to find} the good parameters: the existence of an excellent approximator $\theta^*$ somewhere in parameter space says nothing about whether gradient descent will ever take you there. The latter is a dynamics question, and we will return to it.

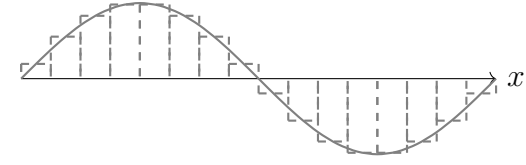
\begin{figure}
    \centering
\begin{tikzpicture}

    \draw[->] (0,0) -- (2*pi,0) node[right] {$x$};

    \draw[domain=0:2*pi, smooth, variable=\x, gray, thick]
        plot ({\x}, {sin(\x r)});

    \foreach \i in {0,...,15}
    {
        \pgfmathsetmacro{\xstart}{\i * pi / 8}
        \pgfmathsetmacro{\xend}{(\i + 1) * pi / 8}
        \pgfmathsetmacro{\xmid}{(\xstart + \xend) / 2}
        \pgfmathsetmacro{\ymid}{sin(\xmid r)}
        \draw[gray, dashed, thick] (\xstart,\ymid) -- (\xend,\ymid);
    }

    \foreach \i in {1,...,15}
    {
        \pgfmathsetmacro{\xpos}{\i * pi / 8}
        \pgfmathsetmacro{\xprev}{(\i - 1) * pi / 8}
        \pgfmathsetmacro{\xnext}{\i * pi / 8}
        \pgfmathsetmacro{\ymid}{sin((\xpos - pi / 16) r)}
        \pgfmathsetmacro{\yprev}{sin((\xprev + \xnext) / 2 r)}
        \draw[gray, dashed, thick] (\xpos, \yprev) -- (\xpos,0);
        \draw[gray, dashed, thick] (\xpos - pi/8, \yprev) -- (\xpos - pi/8,0);
    }

    \pgfmathsetmacro{\yinit}{sin((0 + pi/8) / 2 r)}
    \pgfmathsetmacro{\yfinal}{sin((15*pi/8 + 2*pi) / 2 r)}
    \draw[gray, dashed, thick] (0, \yinit) -- (0, 0);
    \draw[gray, dashed, thick] (2*pi, \yfinal) -- (2*pi, 0);

\end{tikzpicture}
\caption{The mechanism behind the Universal Approximation Theorem: tile $\sin(x)$, or anything else, with a sequence of bumps built from differences of (approximate) step functions.}
\label{fig:UAT_sin}l
\end{figure}

To ask the obvious,
\Q{Why does the UAT work?}
Cybenko worked with the sigmoid
\begin{equation}
\sigma(x) = \frac{1}{1+e^{-x}},
\end{equation}
which makes the mechanism easy to draw. Each unit contributes
\begin{equation}
\sigma\left(w_i^{(0)}\cdot x + b_i^{(0)}\right),
\end{equation}
which limits to a shifted step function as $w^{(0)}\to\infty$. Differences of steps make localized bumps, the output weights $w^{(1)}$ scale the bumps, and bumps tile your favorite function; see Fig.~\ref{fig:UAT_sin}. Cybenko's theorem \cite{cybenko1989approximation} was only the beginning; generalizations abound, including to multiple hidden layers \cite{hornik1991approximation}, other activations, and other domains.

\subsubsection{Kolmogorov-Arnold Theorem}
\begin{figure}[th]
    \begin{center}
    \includegraphics[width=1.05\columnwidth]{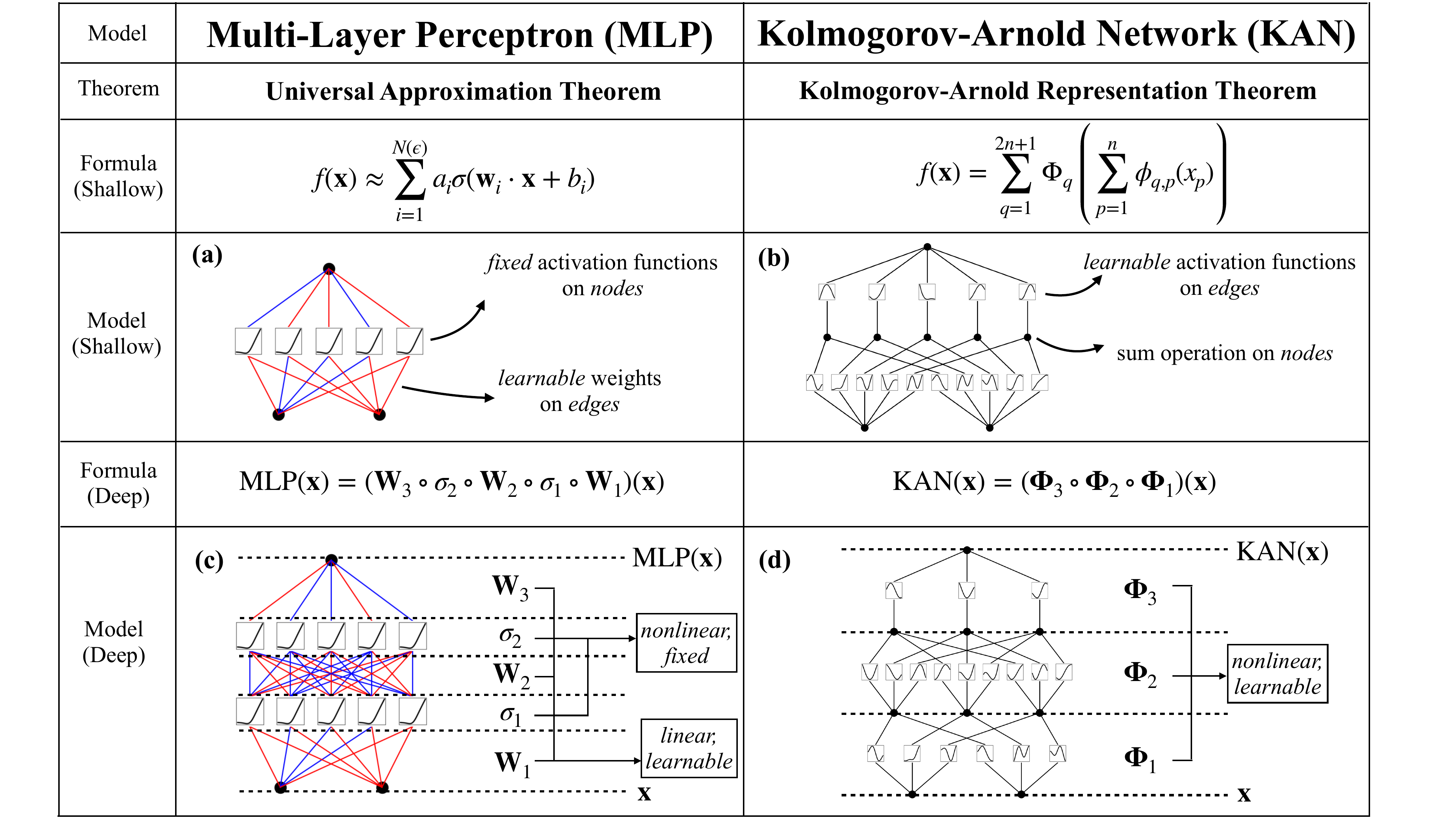}
    \end{center}
    \caption{MLPs versus KANs: functional forms, the mathematical theorems motivating each, and how each is stacked into layers.}
    \label{fig:KAN-figure}
    \end{figure}
The UAT says complicated functions can be assembled from simple building blocks. One naturally wonders what other theorems have this flavor, and whether they, too, suggest architectures.

A beautiful example is due to Kolmogorov \cite{kolmogorov1957representation} and Arnold \cite{arnold1957representation}: every multivariate continuous function can be written --- represented exactly, not merely approximated --- using only one-dimensional functions and addition:
\begin{theorem}[Kolmogorov-Arnold Representation Theorem]
Let $f:[0,1]^n\to \bR$ be an arbitrary multivariate continuous function. Then it has the representation
\begin{equation}
    f(x_1,\dots,x_n) = \sum_{q=0}^{2n}\Phi_q\left(\sum_{p=1}^n \phi_{q,p}(x_p)\right) \label{eqn:KART}
\end{equation}
with continuous one-dimensional functions $\Phi_q$ and $\phi_{q,p}$.
\end{theorem}

At first sight this seems too good: if univariate functions plus addition suffice, what about $f(x,y)=xy$? That particular case is no mystery,
\begin{equation}
    f(x,y) = xy = e^{\log x + \log y},
\end{equation}
but the general statement is highly non-trivial. Alternatively,
\begin{equation}
f(x,y) = xy = \frac{1}{4}(x+y)^2 - \frac{1}{4}(x-y)^2
\end{equation}
if one wishes to work on the whole unit square.

\medskip
Look closely at the structure of \eqref{eqn:KART}: it is a sum of functions of sums of functions, which smells like a neural network. But there is a twist. The inner functions $\phi_{q,p}$ carry both indices $p$ and $q$, so they naturally live on the \emph{edge} connecting input $x_p$ to the intermediate quantity $x^{(1)}_{q} := \sum_{p} \phi_{q,p}(x_p)$, and distinct edges carry independent functions. If this is a network, it is one whose activation functions sit on edges rather than nodes, and are themselves learnable.

That observation became the Kolmogorov-Arnold network (KAN) \cite{KAN}, an architecture with learnable edge activations, stackable into deep layers just as the single-layer UAT structure stacks into MLPs. See Fig.~\ref{fig:KAN-figure} for the comparison. A notable feature of KANs in practice is that the learned univariate activations can be visualized and often snapped to symbolic form, yielding interpretable models that map onto closed-form expressions, a feature we will exploit in Day 3 when we discuss conjecture generation.

\subsubsection{Attention and the Transformer}

The UAT and Kolmogorov-Arnold theorems both take a fixed input vector $x\in\bR^d$ and ask what can be built on top of it. But language, the medium of the reasoning models and agents of the Introduction, does not arrive as a fixed vector; it arrives as a \emph{sequence} of tokens of variable length. So we ask
\Q{What architecture reads a sequence?}

The dominant answer is \textbf{attention} \cite{vaswani2017attention}, and the intuition is linguistic. A \textbf{token} is the atomic unit a model reads: \textbf{tokenization} splits raw text into these units (usually subword fragments chosen by a frequency-based scheme such as byte-pair encoding, so common words stay whole while rare ones break into pieces), each mapped to an integer id and then to a learned embedding vector. Take the tokens to be the words of an English sentence (real systems use subword fragments, but words will do). Each word arrives with a \emph{context-free} embedding: picture its Oxford English Dictionary entry, a fixed vector cataloguing every sense the word can carry. ``Bank'' gets the same entry whether you are standing on a river or in a vault. But one does not read a dictionary; one reads a sentence, and fixes the sense of each word by \emph{looking at its neighbors}: ``river,'' or ``deposit.'' \textbf{Self-attention} is precisely this operation: it turns the context-free entry of each token into a context-dependent representation by letting every token gather information from the rest.

Concretely, write each token vector as $x\in\bR^{d}$, with $d$ the embedding dimension, and stack the $n$ of them as the rows of $X\in\bR^{n\times d}$. Each token issues a \textbf{query} $q=xW_Q$ (what context do I need?), advertises a \textbf{key} $k=xW_K$ (what context do I offer?), and carries a \textbf{value} $v=xW_V$ (the content it passes along), with learnable $W_Q,W_K\in\bR^{d\times d_k}$ and $W_V\in\bR^{d\times d_v}$. Since queries, keys, and values are all built from the same $X$, this worked example is \emph{self-attention}. The softmax-normalized compatibility of queries with keys decides who listens to whom:
\begin{equation}
\begin{aligned}
\text{Attn}(X)&=\underbrace{\text{softmax}\!\left(\frac{QK^{T}}{\sqrt{d_k}}\right)}_{A}\,V,\\
Q&=XW_Q,\quad K=XW_K,\quad V=XW_V.
\end{aligned}
\end{equation}
Here \begin{equation}\mathrm{softmax}(z)_i = \frac{e^{z_i}}{\sum_j e^{z_j}}\end{equation} turns each row of raw query--key scores into a positive, normalized distribution (a probabilistic choice of what to read rather than a deterministic pick), so the attention matrix $A\in\bR^{n\times n}$ has rows that each sum to one. The $i$-th row of $A$ is a  distribution over the tokens that token $i$ attends to: ``bank'' places its weight on ``river.'' 

Running $N$ such maps in parallel gives \textbf{multi-head attention}; the number of heads is the $N$ of the NNGP list in \S\ref{sec:statistics}. Queries and keys necessarily share a dimension, $d_q=d_k$, since they are dotted together; in practice one also takes $d_k=d_v=d/N$, the \textbf{head dimension}, so the $N$ heads concatenate back to the full width $d$. Interleaving multi-head attention with a per-token MLP, residual connections, and normalization gives the \textbf{transformer} block \cite{vaswani2017attention}; stack $L$ of them and you have the architecture beneath every language model, reasoning model, and agent in these lectures. Self-attention is permutation-equivariant by construction; word order is reinjected by adding \textbf{positional encodings} to the inputs: position-dependent vectors (fixed sinusoids of the token index, or learned per-position embeddings) added to the token embeddings so that otherwise-identical tokens at different positions are distinguishable. 

When $Q,K,V$ all come from one sequence this is \textbf{self-attention}; when the queries come from one sequence and the keys and values from another (a decoder attending to an encoder's output), it is \textbf{cross-attention}. The original transformer was an encoder--decoder built for translation \cite{vaswani2017attention}; most of today's language models are \textbf{decoder-only} \cite{GPT3}, a single causal stack predicting each token from those before it. ``Causal'' is enforced by a \textbf{mask}: before the softmax, the attention scores to future positions are set to $-\infty$ (the strictly upper-triangular part of $A$), so token $i$ attends only to tokens $\leq i$ and next-token prediction never peeks ahead. They train by self-supervised next-token prediction, minimizing the cross-entropy of each token given its predecessors over a large corpus, after which post-training (instruction tuning, and RL such as the GRPO of Day~3) shapes their behavior.

\medskip
With this important contextual setup, we should return to expressivity.
Does the transformer hide anything in principle? Like the perceptron, no:
\begin{theorem}[Yun et al. \cite{yun2020transformers}]
With positional encodings, transformers are universal approximators of continuous sequence-to-sequence functions on a compact domain.
\end{theorem}
\noindent This is an analog of a UAT for sequence data.

But the expressivity story has a second chapter. A transformer of fixed depth executes a fixed number of serial computational steps per forward pass, which provably bounds what a single pass can express: inherently sequential problems lie out of reach. Let the model emit intermediate tokens and read them back (a \textbf{chain of thought} \cite{wei2022chain}) and you hand it serial steps on demand: its expressive power now grows with the length of its reasoning \cite{merrill2024chainofthought,feng2023chainofthought}. The reasoning models of the Introduction, post-trained to think before they answer \cite{openai2024o1,deepseek2025r1}, are cashing in on exactly this.
\CT{Reasoning is Expressivity}{thinking in tokens spends test-time computation to buy expressive power.}

We will meet this architecture again: the redundancy $W_Q\to RW_Q,\ W_K\to RW_K$, which leaves $W_Q^T W_K$ fixed, reappears as a gauge symmetry in the anomaly formalism of Day 2 (\S\ref{sec:anomalies}), and transformers reading braid words will classify knots in Day 3 (\S\ref{sec:RL}).

\subsection{Statistics of Neural Networks}
\label{sec:statistics}

Now let us understand neural networks at initialization, although the questions here hold for any fixed-time $t$ with associated $P_t(\theta)$.

 Booting up your laptop once will not do: a single initialization is a single draw $\theta\sim P(\theta)$, hence a single random function. The object of study is the \emph{ensemble}:
 \Q{What characterizes the stats of the NN ensemble?}
The moments, or $n$-pt functions,
\begin{equation}
G^{(n)}(x_1,\dots,x_n) = \bk{\phi(x_1)\dots \phi(x_n)},
\end{equation}
package this information, and they all descend from a partition function
\begin{align}
Z[J] &= \bk{e^{\int d^dx J(x) \phi(x)}} \\
G^{(n)}(x_1,\dots,x_n) &= \left(\frac{\delta}{\delta J(x_1)}\dots \frac{\delta}{\delta J(x_n)} Z[J]\right)\bigg|_{J=0},
\end{align}
with $J(x)$ a source. I have deliberately left the bracket $\bk{\cdot}$ unspecified. Using the expectation from the Setup,
\begin{align}
    \label{eqn:Zparam}
    Z[J] & = \int d\theta \,P(\theta)\, e^{\int d^dx  J(x) \phi(x)},
\end{align}
an integral over the density of network parameters, recalling that $\phi(x)$ depends on $\theta$. But physicists are accustomed to a different representation of the same kind of object,
\begin{align}
    \label{eqn:Zfunc}
Z[J] & = \int \cD\phi \,e^{-S[\phi]} e^{\int J(x) \phi(x)},
\end{align}
the Feynman path integral, where the density on functions comes from an action $S[\phi]$.

Specifying a neural network means specifying the pair $(\phi,P(\theta))$, so the \emph{parameter-space partition function} \eqref{eqn:Zparam} is natural to study and correlators can always be computed from it. Given that data, it is natural to ask
\Q{What is the action $S[\phi]$ associated to $(\phi,P(\theta))$?}
Whenever this question has an answer, the same theory has two presentations, and the parameter-space and function-space descriptions might deserve to be called a \emph{duality}.

\subsubsection{NNGP Correspondence}

With the question of $S[\phi]$ on the table, we turn to a classic result of Neal \cite{neal} from the 90's.

Take the single-layer fully connected network of width $N$, biases off to reduce clutter:
\begin{equation}
    \phi(x) = \frac{1}{\sqrt{N}}\sum_{i=1}^N  w_i^{(1)} \sigma\left(\sum_{j=1}^d w_{ij}^{(0)} x_j\right), \label{eqn:biasless_perceptron}
    \end{equation}
with parameters $\theta = \{w_{ij}^{(0)}, w_i^{(1)}\}$ drawn independently and identically distributed (i.i.d.),
\begin{equation}
    w_{ij}^{(0)}\sim P(w^{(0)}) \qquad w_i^{(1)} \sim P(w^{(1)}).
\end{equation}
Stare at \eqref{eqn:biasless_perceptron} and notice:
\O{The network is a sum of $N$ i.i.d. functions.}
This is the Central Limit Theorem in function space (see Appendix \ref{app:CLT} for the scalar version), and it yields the Neural Network / Gaussian Process (NNGP) correspondence:
\CT{NNGP Correspondence}{in the $N \to \infty$ limit, $\phi$ is drawn from a Gaussian Process (GP),
\begin{equation}
    \lim_{N\to \infty} \,\,\,\phi(x) \sim \cN\left(\mu(x), K(x,y)\right),
\end{equation}
with mean $\mu(x)$ and covariance (kernel) $K(x,y)$.}
By the CLT, $\exp(-S[\phi])$ becomes Gaussian, i.e. $S[\phi]$ is quadratic in the network. This should set off physics alarm bells: the infinite-width network is a draw from a Gaussian density on functions, which is to say a generalized free field theory.

Before discussing how general this is, let us compute. Take $P(w^{(1)})$ zero-mean with finite variance,
\begin{equation}
    \bk{w^{(1)}}=0 \qquad  \bk{w^{(1)}w^{(1)}} = \mu_2,
\end{equation}
so the one-point function vanishes, $G^{(1)}(x) = 0$. Following Williams \cite{williams}, the two-point function is computed directly in parameter space (Einstein summation):
\begin{align}
    G^{(2)}(x,y)
        & = \frac{1}{N}\bk{w_i^{(1)} \sigma(w_{ij}^{(0)} x_j)\,\,w_k^{(1)}  \sigma(w_{kl}^{(0)} y_l)} \\
        & \teq{} \frac{1}{N}\, \bk{w_i^{(1)} w_k^{(1)}} \bk{\sigma(w_{ij}^{(0)} x_j) \sigma(w_{kl}^{(0)} y_l)} \\
        & = \frac{\mu_2}{N} \bk{\sigma(w_{ij}^{(0)} x_j)\,  \sigma(w_{il}^{(0)} y_l)},
\end{align}
using $\bk{w_i^{(1)}w_k^{(1)}} = \mu_2 \delta_{ik}$ from i.i.d.-ness. The remaining $i$-sum runs over $N$ identical copies, so
\begin{equation}
    G^{(2)}(x,y) = \mu_2\,\, \bk{\sigma(w_{ij}^{(0)} x_j)\,  \sigma(w_{il}^{(0)} y_l)},
\end{equation}
now with \emph{no sum on $i$}. This is exact in $N$, and all that remains is the bra-ket, a finite-dimensional integral over $w^{(0)}$. Do it analytically if you can; if not, Monte Carlo with $M$ samples $w^{(0)}\sim P(w^{(0)})$ gives
\begin{equation}
    G^{(2)}(x,y) \simeq \frac{\mu_2}{M}\sum_{\text{samples}}^M \sigma(w_{ij}^{(0)} x_j)\,  \sigma(w_{il}^{(0)} y_l),
\end{equation}
and standard NN parameter densities are chosen to be trivially sampleable. (For densities that are harder to sample one may use Markov Chain Monte Carlo, as in lattice field theory.)

This pins down the NNGP,
\begin{equation}
    \lim_{N\to \infty} \,\,\,\phi(x) \sim \cN\left(0, G^{(2)}(x,y)\right),
\end{equation}
with action
\begin{equation}
S[\phi] = \frac{1}{2}\int d^dx \,d^dy\, \phi(x) \,G^{(2)}(x,y)^{-1}\, \phi(y),
\end{equation}
where the inverse is defined by
\begin{equation}
\int d^dy \, G^{(2)}(x,y)^{-1} G^{(2)}(y,z) = \delta^{(d)}(x-z).
\end{equation}
So for any NNGP with vanishing mean: compute $G^{(2)}$ in parameter space, invert, and you have the action.

\vspace{.5cm}
Certain big neural networks, then, are draws from generalized free field theories. You should be wondering
\Q{How general is the NNGP correspondence?}
Neal proved it for single-layer feedforward networks, and the result sat quietly for years; my guess is the AI Winter and the field's focus on non-NN methods is the reason. Once NNs took off in the 2010's post-AlexNet \cite{NIPS2012_c399862d}, people asked the question architecture by architecture: does architecture $X$ have a structural hyperparameter $N$ whose $N \to \infty$ limit is a GP? Before reading the list, ask yourself
\Q{Neal's result was just a CLT applied to a sum of $N$ i.i.d. functions. Does this occur often in NNs?}
It does, and accordingly the NNGP limit is everywhere:
\begin{itemize}
    \item \textbf{Deep Fully Connected Networks,} $N=\text{width}$,
    \item \textbf{Convolutional Neural Networks,} $N=\text{channels}$,
    \item \textbf{Attention Networks,} $N=\text{heads}$,
\end{itemize}
and more; see \cite{yang2019wide} and references therein for a unified treatment.

\subsubsection{Non-Gaussian Processes}

If Gaussianity comes from the CLT, then interactions (non-Gaussianities, in field theory language) come from violating CLT assumptions. From Appendix \ref{app:CLT}, the two levers are finite-$N$ corrections and breaking statistical independence; \cite{Demirtas:2023fir} develops independence-breaking systematically, including the extraction of NN actions from correlators.

Let us exhibit the $N$-dependence of the connected four-point function. Williams' trick works for any correlator, and to spare ourselves index gymnastics we write the architecture as
\begin{equation}
\phi(x) = \frac{1}{\sqrt{N}} \sum_i w_i \varphi_i(x),
\end{equation}
with $w_i$ distributed like the output weights above and $\varphi_i(x)$ i.i.d. neurons of \emph{arbitrary} internal architecture. Then
\begin{align}
G^{(4)} &= \bk{\phi(x)\phi(y)\phi(z)\phi(w)} \\
    &= \frac{1}{N^2}\sum_{i,j,k,l} \bk{w_i w_j w_k w_l} \bk{\varphi_i(x)\varphi_j(y)\varphi_k(z)\varphi_l(w)} \\
    &= \frac{1}{N^2}\sum_i \bk{w_i^4} \bk{\varphi_i(x)\varphi_i(y)\varphi_i(z)\varphi_i(w)} \\ &\,\,\,\,\,\,\,\,+\frac{1}{N^2} \sum_{i\neq j} \bk{w_i^2} \bk{w_j^2}\bk{\varphi_i(x)\varphi_i(y)\varphi_j(z)\varphi_j(w) + \text{perms}}.
\end{align}
Watch the indices: they are doing a lot of work. The connected four-point function is
\begin{equation}
G^{(4)}_c(x,y,z,w) = G^{(4)}(x,y,z,w) - \left(G^{(2)}(x,y)G^{(2)}(z,w) + \text{perms}\right)
\end{equation}
comes out to
\begin{align}
G^{(4)}_c(x,y,z,w) = \frac{1}{N} \bigg(\mu_4 \, \bbk{\varphi_i(x)\varphi_i(y)\varphi_i(z)\varphi_i(w)} \\ - \mu_2^2 \,\left(\bbk{\varphi_i(x)\varphi_i(y)}\bk{\varphi_i(z)\varphi_i(w)} + \text{perms}\right)\bigg),
\end{align}
no Einstein summation on $i$. The connected four-point function is $O(1/N)$: finite width means interactions, with strength controlled by $1/N$. In examples below, $G^{(4)}_c$ is computable in closed form.

The connected correlators exhibit CLT-scaling with $N$ more generally. Specifically, 
\begin{equation}
G^{(2n)}_c \propto \frac{1}{N^{n-1}} \qquad n\geq 2,
\end{equation}
which is why the higher connected correlators vanish in the $N\to\infty$ limit, leaving only the two-point function and Gaussianity.

\subsubsection{Symmetries}
We have a handle on the statistics. The next question is
\Q{Does the \emph{ensemble} have structure?}
That is, properties of the ensemble that no individual network knows about? A  broad version of this question is the subject of Day 2; here we focus on symmetries.

To accommodate symmetries acting at input and output, consider networks
\begin{equation}
\phi: \bR^d \to \bR^D
\end{equation}
with $D$-dimensional output, indices sometimes suppressed.

A classic notion at the level of \emph{individual} networks is equivariance \cite{pmlr-v48-cohenc16}. A network is \textbf{$G$-equivariant} for a group $G$ if
\begin{equation}
\rho_{D}(g) \phi(x) = \phi(\rho_d(g) x) \qquad \forall g \in G,
\end{equation}
with $\rho_d, \rho_D$ matrix representations of $G$ on input and output space, and \textbf{invariant} when $\rho_D = \mathbbm{1}$. Building the right equivariance into an architecture can dramatically improve learning speed and sample efficiency \cite{winkels20183d}, even at the level of scaling laws \cite{kaplan2020scaling, Batzner2022,frey2022neural}; lattice field theorists will recognize $SU(N)$-equivariant flows \cite{boyda2021sampling} as a prominent physics example.

But this is the statistics lecture, so we want symmetries of \emph{ensembles}: transformations that leave the statistical ensemble invariant. In field theory these are \textbf{global symmetries}. Let the network transform as
\begin{equation}
\phi \mapsto \phi_g, \qquad g \in G.
\end{equation}
The ensemble has global symmetry group $G$ if the partition function is invariant,
\begin{equation}
    Z_g[J] = Z[J] \qquad \forall g \in G,
\end{equation}
i.e.
\begin{equation}
 \bk{e^{\int d^d x J(x) \phi_g(x)}}  =\bk{e^{\int d^d x J(x) \phi(x)}}  \qquad \forall g \in G,
\end{equation}
with indices on $\phi$ and $J$ as needed. After a network redefinition on the LHS, the statement becomes invariance of $\bk{\cdot}$ itself. In the path integral this is the familiar invariance of $S[\phi]$ and $\cD\phi$. In parameter space, the mechanism \cite{Maiti:2021fpy} is to absorb $g$ into a reparameterization $\theta \mapsto \theta_g$, and the symmetry holds when
\begin{equation}
\int d\theta_g \,P(\theta_g)\, e^{\int d^d x J(x)\phi_\theta(x)} = \int d\theta\, P(\theta) \,e^{\int d^d x J(x)\phi_\theta(x)},
\end{equation}
i.e. when the parameter density and measure are invariant under the absorbed action. Keep this absorption mechanism in mind: in Day 2 it drives our analysis of Ward identities and anomalies.

Transforming inputs or outputs gives analogs of spacetime and internal symmetries, respectively; transformations of intermediate layers are also conceivable if one wants symmetric learned representations. Equivariance plugs in naturally, converting an input transformation into an output transformation, whose induced action must then preserve the partition function.

\medskip
\noindent \textbf{Example.} \label{sec:stats_examples} For any architecture of the form
\begin{equation}
    \phi(x) = \sum_i w_i \varphi_i(x), \qquad w_i \sim P(w) \, \text{even},
\end{equation}
with arbitrary neuron $\varphi$ (deep, shallow, whatever), the $\bZ_2$ action
$\phi \mapsto -\phi$ is absorbed by $w_g = -w$, and evenness of $P(w)$ makes $dw\, P(w)$ invariant. Provided the domain is invariant this is a global symmetry, and all odd correlators vanish.

\subsubsection{Examples}

We have collected three statistical phenomena: Gaussian limits, non-Gaussianities from CLT violation, and symmetries. Let us see them all concretely in a few architectures. For standard ML cases such as ReLU networks, see \cite{Halverson:2020trp} and references therein.

\medskip
\noindent \textbf{Gauss-net.}
Architecture and densities:
\begin{align}
\phi(x) = \frac{w^{(1)}_{i} \exp(w^{(0)}_{ij} x_{j} + b^{(0)}_{i})}{\sqrt{\exp[2(\sigma_{b_0}^2 + \frac{\sigma^2_{w_0}}{\din}x^2 )]}},
\end{align}
with i.i.d. draws
\begin{equation}
    w^{(0)} \sim \mathcal{N}(0,\frac{ \sigma^2_{W_0}}{\din}),\qquad w^{(1)} \sim \mathcal{N}(0, \frac{\sigma^2_{w_1}}{2N}),\qquad b^{(0)} \sim \mathcal{N}(0, \sigma^2_{b_0}).
\end{equation}
The two-point function is
\begin{equation}
    G^{(2)}(x_1, x_2) = \frac{\sigma_{w_1}^2}{2} e^{-\frac{1}{2\din}\sigma_{w_0}^{2} (x_1-x_2)^2},
\end{equation}
exhibiting a correlation length
\begin{equation}
\xi = \sqrt{\frac{\din}{\sigma_{w_0}^2}}.
\end{equation}
The connected four-point function is
\begin{align}
    G^{(4)}_c =&\frac{1}{4 N} \sigma_{w_1}^4 \bigg[ 3e^{4\sigma_{b_0}^2} e^{-\frac{\sigma_{w_0}^2}{2d}\sum_i x_i^2 -2(x_1 x_2 + \,6 \text{ perms})} \nonumber \\ &- (e^{-\frac{1}{2d} \sigma_{w_0}^2
    \left((x_{1}-x_{4})^2+(x_{2}-x_{3})^2\right)} + \,2 \text{ perms}
    ) \bigg].
\end{align}
Note the asymmetry: the $2$-pt function is translation invariant, the $4$-pt function is not.

\medskip
\noindent \textbf{Euclidean-Invariant Nets}.
Suppose we want invariance under the full Euclidean group ($SO(d)$ rotations and translations) built in at the level of the input layer. Consider
\begin{equation}
\ell_i(x) = F({\bf w^{(0)}}_i) \cos\left(\sum_j w^{(0)}_{ij} x_j + b^{(0)}_i\right), \,\,\, i \in {1,\dots,N},
\end{equation}
sums explicit since $i$ is not summed, with
\begin{equation}
    w^{(0)}_{ij} \sim P(w^{(0)}_{ij}) \qquad b^{(0)}_i \sim \text{Unif}[-\pi,\pi].
\end{equation}
A rotation $x_i\to R_{ij} x_j$, $R\in SO(d)$, is absorbed into a redefinition of $w^{(0)}_{ij}$, leaving the partition function invariant provided $F({\bf w^{(0)}}_i)$ and $P(w^{(0)}_{ij})$ are rotation-invariant. A translation $x_j \to x_j+\epsilon_j$ produces $w^{(0)}_{ij}\epsilon_j$, absorbed into the uniform phase $b^{(0)}_i$. Two features deserve emphasis:
\begin{itemize}
\item \textbf{Larger Euclidean Nets.} Any network built on top of $\ell$ that does not reuse its parameters inherits Euclidean invariance.
\item \textbf{Spectrum Shaping.} In $G^{(2)}(p)$, the function $F$ gets evaluated on momenta, and may be chosen to sculpt the power spectrum (momentum-space propagator) at will.
\end{itemize}
General correlators are in \cite{Halverson:2021aot}; remember spectrum shaping, since Day 2 leans on it hard. A scalar NN follows by contracting with output weights,
\begin{equation}
    \label{eqn:single_layer_Euclidean}
    \phi(x) = \sum w^{(1)}_i \ell_i(x), \qquad w^{(1)}_i \sim P(w^{(1)}).
\end{equation}

\medskip
\noindent \textbf{Cos-net.}
Specialize \eqref{eqn:single_layer_Euclidean} to $F=1$,
\begin{equation}
\phi(x) =  w^{(1)}_i \cos\left( w^{(0)}_{ij} x_j + b^{(0)}_i\right),
\end{equation}
with densities
\begin{equation}
    w^{(1)} \sim \cN\left(0,\frac{\sigma_{w_1}^2}{N}\right) \qquad w^{(0)} \sim \cN\left(0,\frac{\sigma_{w_0}^2}{d}\right) \qquad b^{(0)} \sim \text{Unif}[-\pi,\pi].
\end{equation}
The two-point function coincides with Gauss-net's,
\begin{equation}
    G^{(2)}(x,y) = \frac{\sigma_{w_1}^2}{2} e^{-\frac{1}{2\din}\sigma_{w_0}^{2} (x-y)^2},
\end{equation}
while the connected four-point function is
\begin{align}
    G^{(4)}|_{c} = &\frac{1}{8 N} \sigma_{w_1}^4 \bigg[ 3  \left(e^{-\frac{1}{2d} \sigma_{w_0}^2
   (x_{1}+x_{2}-x_{3}-x_{4})^2}+\, 2\text{ perms}\right) \nonumber \\
   &-\left(2 e^{-\frac{1}{2d} \sigma_{w_0}^2
   \left((x_{1}-x_{4})^2+(x_{2}-x_{3})^2\right)}+\, 2\text{ perms}\right) \bigg],
\end{align}
fully Euclidean invariant. Comparing the two architectures:
\begin{itemize}
\item \textbf{Symmetry.} Cos-net is Euclidean-invariant at \emph{every} $N$, by construction; Gauss-net only as $N\to\infty$, where the theory is determined by $G^{(2)}$ alone.
\item \textbf{Large-$N$ Duality.} As $N\to\infty$ the two theories converge to the \emph{same} Gaussian Process.
\end{itemize}

\medskip
\noindent \textbf{Scalar-net.}
One last example, chosen to reproduce the free scalar. Specialize \eqref{eqn:single_layer_Euclidean} to
\begin{equation}
    \label{eqn:scalar-net}
    \phi(x) =  \sqrt{\frac{2 \text{vol}(B_\Lambda^d)}{(2\pi)^d \sigma_{w_1}^2}} \frac{1}{\sqrt{{\bf w^{(0)}}_i^2 + m^2}} \,\,w^{(1)}_i \cos\left( w^{(0)}_{ij} x_j + b^{(0)}_i\right),
\end{equation}
with densities
\begin{equation}
    \label{eqn:scalar-net-densities}
    w^{(1)} \sim \cN\left(0,\frac{\sigma_{w_1}^2}{N}\right) \qquad w^{(0)} \sim \text{Unif}(B_\Lambda^d) \qquad b^{(0)} \sim \text{Unif}[-\pi,\pi],
\end{equation}
where $B_\Lambda^d$ is the radius-$\Lambda$ ball in momentum space. Translation invariance is built in, and the power spectrum evaluates to
\begin{equation}
G^{(2)}(p) = \frac{1}{p^2+m^2}.
\end{equation}
This is the free scalar in $d$ Euclidean dimensions, realized as an ensemble of neural networks. Hold that thought until Day 2, where we deform it to $\phi^4$ and, with relatives of this architecture, to strings.

\subsection{Dynamics of Neural Networks}
\label{sec:dynamics}

Two pillars down, one to go:
\Q{How does a NN evolve under gradient descent?}
The plan: first a simplification known as the neural tangent kernel (NTK), then an exactly solvable model with MSE loss, then a critique of the NTK regime, and finally a scaling analysis that fixes its central deficiency and gives feature learning.

The setting is supervised learning with gradient descent: data
\begin{equation}
\cD = \{(x_\alpha,y_\alpha)\}_{\alpha=1}^{|\cD|},
\end{equation}
loss
\begin{equation}
    \cL[\phi] = \frac{1}{|\cD|}\sum_{\alpha=1}^{|\cD|} \ell(\phi(x_\alpha), y_\alpha),
\end{equation}
and parameter updates
\begin{equation}
    \frac{d\theta_i}{dt} = -\eta \nabla_{\theta_i} \cL[\phi].
\end{equation}
Defining
\begin{equation}
\Delta(x) = - \frac{\delta\ell(\phi(x),y)}{\delta \phi(x)},
\end{equation}
with $y$ the label attached to $x$, the chain rule recasts gradient descent as
\begin{equation}
\label{eqn:gd_with_delta}
\frac{d\theta_i}{dt}= \frac{\eta}{|\cD|}\sum_{\alpha=1}^{|\cD|} \Delta(x_\alpha) \frac{\partial \phi(x_\alpha)}{\partial \theta_i}.
\end{equation}
$\Delta(x)$ is the natural function-space residue of the loss.

\textbf{Einstein summation is in force throughout this section unless suspended explicitly (which will happen).}

\subsubsection{Neural Tangent Kernel}

A classic of ML theory \cite{NTK}. Push the dynamics from parameters to functions:
\begin{align}
\frac{d\phi(x)}{dt} &= \frac{\partial\phi(x)}{\partial\theta_i} \frac{d\theta_i}{dt} \\ &= \frac{\eta}{|\cD|}\sum_{\alpha=1}^{|\cD|} \Delta(x_\alpha) \,\Theta(x,x_\alpha),
\end{align}
where
\begin{equation}
    \Theta(x,x_\alpha) = \frac{\partial\phi(x)}{\partial\theta_i}  \frac{\partial\phi(x_\alpha)}{\partial\theta_i}
\end{equation}
is the \emph{neural tangent kernel} (NTK).

The derivation took three lines, so the NTK is clearly fundamental, and yet it looks hopeless to work with, being:
\begin{itemize}
\item \textbf{Parameter-dependent.} It has a sum over parameters, which number in the trillions for the strongest modern NNs.
\item \textbf{Time-dependent.} It depends on parameters, which time evolve along the trajectory $\theta(t)$.
\item \textbf{Stochastic.} It inherits randomness from $\theta(0) \sim P(\theta)$.
\end{itemize}
It is also non-local, carrying loss information from train points $x_\alpha$ to the test point $x$. This is an unwieldy object.

What earns the NTK its fame is the $N\to \infty$ limit. There, training enters the
\begin{equation}
\text{Lazy regime:} \qquad |\theta(t) - \theta(0)| \ll 1,
\end{equation}
where parameters are so numerous that each barely moves, and the network is well-approximated by its linearization \cite{NTK,lee2019wide}
\begin{equation}
\lim_{N\to \infty} \phi(x)\simeq \phi_\text{lin}(x) := \phi_{\theta_0}(x) + (\theta-\theta_0)_i \frac{\partial\phi(x)}{\partial\theta_i}\bigg|_{\theta_0},
\end{equation}
giving
\begin{equation}
    \lim_{N\to \infty}\Theta(x,x') \simeq \Theta(x,x')\big|_{\theta_0}:
\end{equation}
the infinite-width NTK is the NTK at initialization. Better still, in the same limit the law of large numbers converts sums into expectations,
\begin{equation}
\lim_{N\to \infty} \Theta(x,x')|_{\theta_0} = \bk{\beta_\theta(x,x')} =: \bar \Theta(x,x'),
\end{equation}
for a computable $\beta(x,x')$, and the dynamics collapses to
\begin{equation}
    \label{eqn:dynamics_frozen_NTK}
    \frac{d\phi(x)}{dt}= -\frac{\eta}{|\cD|}\sum_{\alpha=1}^{|\cD|} \frac{\delta \ell(\phi(x_\alpha),y_\alpha)}{\delta\phi(x_\alpha)} \,\, \bar \Theta(x,x_\alpha),
\end{equation}
governed by the \emph{frozen NTK} $\bar \Theta$: deterministic, computable at initialization, fixed once and for all. An enormous simplification.

And yet, you should complain!
\CT{Complaint}{The dynamics \eqref{eqn:dynamics_frozen_NTK} communicates loss information from train points $x_\alpha$ to test points $x$ through a \emph{fixed} kernel.}
There are \emph{zero} parameters in sight, and since the network only enters through the frozen $\bar\Theta$, nothing happening inside the network during training affects the evolution. In this limit the NN \textbf{does not learn features}: the hidden-layer representations stay in a small neighborhood of initialization. This cannot be the crux of deep learning.

\medskip
\noindent \textbf{Example.}
The frozen NTK of the single-layer network, to see the mechanics. With
\begin{equation}
    \phi(x) = \frac{1}{\sqrt{N}} \sum_{i=1}^N  w_i^{(1)} \sigma\left(\sum_{j=1}^d w_{ij}^{(0)} x_j\right)
\end{equation}
(sums explicit; one of them matters), the NTK is
\begin{align}
\Theta(x,x') &= \sum_{i} \frac{\partial \phi(x)}{\partial w_i^{(1)}} \frac{\partial \phi(x')}{\partial w_i^{(1)}} + \sum_{ij} \frac{\partial \phi(x)}{\partial w_{ij}^{(0)}} \frac{\partial \phi(x')}{\partial w_{ij}^{(0)}} \\
&= \frac{1}{N} \sum_{i=1}^N \bigg(\sum_{j,l=1}^d \sigma(w_{ij}^{(0)}x_j)\sigma(w_{il}^{(0)}x'_l)\\ &\qquad + \sum_{j=1}^d x_j x'_j \,w_i^{(1)}w_i^{(1)}\, \sigma'(w_{ij}^{(0)} x_j) \sigma'(w_{ij}^{(0)}x'_j) \bigg) \\ &=: \frac{1}{N} \sum_{i }\beta_i(x,x').
\end{align}
Squint: the $i$-sum runs over $N$ i.i.d. copies of one object $\beta_i(x,x')$, randomness coming from the $i$-direction parameter draws. The law of large numbers then gives, at $N \to \infty$,
\begin{equation}
\bar \Theta(x,x') = \bk{\beta_i(x,x')},
\end{equation}
no sum on $i$. We emphasize
\O{The $N\to \infty$ NTK is deterministic: it depends only on $P(\theta)$, not on the draw.}
When the expectation is doable in closed form, the kernel governing the dynamics is known exactly, forever.

\subsubsection{An Exactly Solvable Model}

Specialize frozen-NTK dynamics to MSE loss,
\begin{equation}
    \ell(\phi(x),y) = \frac{1}{2}(\phi(x)-y)^2,
\end{equation}
so that \eqref{eqn:dynamics_frozen_NTK} becomes
\begin{equation}
    \frac{d\phi(x)}{dt} = -\frac{\eta}{|\cD|}\sum_{\alpha=1}^{|\cD|} (\phi(x_\alpha)-y_\alpha) \bar \Theta(x,x_\alpha).
\end{equation}
This linear ODE solves exactly:
\begin{equation}
    \phi_t(x) = \phi_0(x) + \bar \Theta(x,x_\alpha) \bar\Theta(x_\alpha, x_\beta)^{-1}\left(\mathbbm{1}-e^{-\frac{\eta}{|\cD|} \bar\Theta t}\right)_{\beta \gamma}\left(y_\gamma-\phi_0(x_\gamma)\right),
\end{equation}
the only computational expense being the inversion of the $|\cD|\times|\cD|$ matrix $\bar\Theta(x_\alpha, x_\beta)$, at $O(|\cD|^3)$. The solution traces a path through function space from $\phi_0$ to
\begin{equation}
    \phi_\infty(x) = \phi_0(x) +\bar \Theta(x,x_\alpha) \bar\Theta(x_\alpha, x_\beta)^{-1}\left(y_\beta-\phi_0(x_\beta)\right).
\end{equation}
This is $\textbf{kernel regression}$, an old workhorse of ML. Normally one picks the kernel by hand; here, GD training at $N\to\infty$ \emph{is} kernel regression with a kernel dictated by the architecture, namely $\bar\Theta$.

On train points the network achieves \textbf{memorization},
\begin{equation}
    \phi_\infty(x_\alpha) = y_\alpha \qquad \forall \alpha.
\end{equation}
On test points it interpolates, propagating train-point residuals through the fixed kernel. Averaging over initializations with $\bk{\phi_0}=0$ (true of standard initializations),
\begin{equation}
\mu_\infty(x) := \bk{\phi_\infty(x)} = \bar\Theta(x,x_\alpha) \bar\Theta(x_\alpha, x_\beta)^{-1}y_\beta.
\end{equation}
Let us put some English on the \textbf{remarkable facts}:
\begin{itemize}
    \item $\mu_\infty(x)$ is the mean prediction of infinitely many infinitely wide networks trained for infinite time: the answer to the question we posed in the Setup.
    \item  If $\phi_0$ is a GP draw, so is $\phi_\infty$, with mean $\mu_\infty(x)$; see \cite{lee2019wide} for the covariance.
\end{itemize}

\subsubsection{Feature Learning}

The frozen NTK is a beautiful, tractable toy, but with no feature learning. In this section we do a thorough scaling analysis, asking how to choose the $N$-dependence of the architecture and learning rate such that features evolve non-trivially under gradient descent. The targets: features finite at initialization, predictions evolving in finite time, features evolving in finite time.

Fair warning: this is the densest section of Day 1. If the indices start to swim, hold onto the three goals; everything below is bookkeeping in their service. I follow the lecture notes of Pehlevan and Bordelon \cite{pehlevan2024lecture} with notation adjusted to match these lectures; original literature includes \cite{yang2020feature,bordelon2022self}, with related approaches in \cite{roberts2022principles,yaida2022meta}.

Take a depth-$L$ feedforward network of width $N$,
\begin{equation}
\phi:\bR^D \to \bR
\end{equation}
(input dimension $D$ now, since $d$ is spoken for), defined recursively by
\begin{align}
\phi(x) &= \frac{1}{\gamma_0 N^d}z^{(L)}(x) \\
z^{(L)}(x) &= \frac{1}{N^{a_L}} w_i^{(L)}\, \sigma(z_i^{(L-1)}(x))\\
z^{(\ell)}_i(x) &= \frac{1}{N^{a_\ell}} W^{(\ell)}_{ij} \sigma(z_j^{(\ell-1)}(x)) \\
z_i^{(1)}(x) &= \frac{1}{N^{a_1}\sqrt{D}} W^{(1)}_{ij}x_j,
\end{align}
Einstein summation throughout, Latin indices in $\{1,\dots,N\}$ \emph{except} the first-layer $j \in \{1,\dots,D\}$. Parameters are drawn
\begin{equation}
w_i^{(\ell)} \sim \cN\left(0,\frac{1}{N^{b_L}}\right) \qquad W^{(\ell)}_{ij} \sim \cN\left(0,\frac{1}{N^{b_\ell}}\right),
\end{equation}
and the learning rate is scaled as
\begin{equation}
\eta = \eta_0 \gamma_0^2 N^{2d-c}
\end{equation}
with $\gamma_0,\eta_0$ of order one. The $z$'s are the pre-activations (the features), and Greek subscripts abbreviate inputs, $z^{(\ell)}_\alpha:= z^{(\ell)}(x_\alpha)$.

This is a standard MLP whose ignorance about $N$-scaling has been honestly parameterized by $(a_\ell,b_\ell,c,d)$. Now impose the \textbf{goals}:
\begin{itemize}
    \item \textbf{Finite Initialization Pre-activations.} $z^{(\ell)} \sim O_N(1)\,\,\,\forall \ell$.
    \item \textbf{Learning in Finite Time.} $d\phi(x)/dt\sim O_N(1)$.
    \item \textbf{Feature Learning in Finite Time.} $dz^{(\ell)}/dt \sim O_N(1) \,\,\, \forall \ell$.
\end{itemize}
These constraints admit a one-parameter family of solutions, pinned down completely by one further demand on the learning rate. One constraint at a time.

\bigskip
\noindent \textbf{Finite Pre-activations.} Zero-mean weights give
\begin{equation}
\bk{z^{(\ell)}_\alpha} = 0 \qquad \forall \ell,
\end{equation}
trivially, so the content is in the covariance. First layer:
\begin{align}
\bk{z_{i\alpha}^{(1)}z_{j\beta}^{(1)}} = \frac{1}{D N^{2a_1}} \bk{W^{(1)}_{im}W^{(1)}_{jn}} \, x_{m\alpha}x_{n\beta} = \delta_{ij}\frac{1}{DN^{2a_1+b_1}} x_{m\alpha}x_{m\beta}.
\end{align}
Higher layers:
\begin{align}
    \bk{z_{i\alpha}^{(\ell)}z_{j\beta}^{(\ell)}} &= \frac{1}{ N^{2a_\ell}} \bk{W^{(\ell)}_{im}W^{(\ell)}_{jn}} \, \bk{\sigma(z^{(\ell-1)}_{m\alpha})\sigma(z^{(\ell-1)}_{m\beta})} \\ &= \delta_{ij}\frac{1}{N^{2a_\ell+b_\ell-1}} \frac{1}{N} \bk{\sigma(z^{(\ell-1)}_{m\alpha})\sigma(z^{(\ell-1)}_{m\beta})} \\
    &= \delta_{ij}\frac{1}{N^{2a_\ell+b_\ell-1}} \bk{\Phi^{(\ell-1)}_{\alpha\beta}},
\end{align}
where
\begin{equation}
    \Phi^{(\ell)}_{\alpha\beta} := \frac{1}{N} \sigma(z^{(\ell)}_{m\alpha})\sigma(z^{(\ell)}_{m\beta})
\end{equation}
is the \emph{feature kernel}. Proceeding by induction (assume $z^{(\ell-1)} \sim O_N(1)$, so $\Phi^{(\ell-1)} \sim O_N(1)$ as an average of $N$ order-one terms), finiteness of all pre-activations requires
\begin{equation}
2a_1+b_1=0 \qquad 2a_\ell + b_\ell = 1 \,\, \forall \ell > 1,
\end{equation}
as is empirically necessary for trainable networks.

A worthwhile aside: at $N\to \infty$ the feature kernels become deterministic (echoing the frozen NTK), intermediate pre-activations become Gaussian, and the statistics of a random deep network is a tower of generalized free field theories with correlations propagating layer to layer by recursion \cite{schoenholz2016deep,lee2019wide}.

\bigskip
\noindent \textbf{Learning in Finite Time.} As in the NTK derivation,
\begin{equation}
\frac{d\phi(x)}{dt} = \frac{\eta}{|\cD|}\sum_{\alpha=1}^{|\cD|} \Delta(x_\alpha) \frac{\partial\phi(x_\alpha)}{\partial\theta_i}\frac{\partial\phi(x)}{\partial\theta_i},
\end{equation}
so
\begin{equation}
\frac{d\phi(x)}{dt}\sim O_N(1)\,\,\, \leftrightarrow \,\,\,\frac{\gamma_0^2 N^{2d}}{N^c} \frac{\partial\phi(x_\alpha)}{\partial\theta_i}\frac{\partial\phi(x)}{\partial\theta_i} \sim O_N(1).
\end{equation}
A chain-rule computation through the layers (straightforward, layered, tedious) gives
\begin{align}
    \frac{\eta_0\gamma_0^2 N^{2d}}{N^c} \frac{\partial\phi(x_\alpha)}{\partial\theta_i}\frac{\partial\phi(x)}{\partial\theta_i}
&= \frac{1}{N^c} \bigg[ \frac{1}{N^{2 a_L - 1}} \Phi_{\mu \nu}^{(L-1)} \nonumber \\ &+ \sum_{\ell=2}^{L-1} \frac{1}{N^{2 a_\ell - 1}} G_{\alpha x}^{(\ell)} \Phi_{\alpha x}^{(\ell-1)} + \frac{1}{N^{2 a_1}} G_{\alpha x}^{(1)} \Phi_{\alpha x}^{(0)} \bigg]
\end{align}
where
\begin{equation}
G_{\mu\nu}^{(\ell)} = \frac{1}{N} g^{(\ell)}_{i\mu}\, g^{(\ell)}_{i\nu},
 \qquad g^{(\ell)}_{i\mu} = \sqrt{N}\, \frac{\partial z^{(L)}_\mu}{\partial z^{(\ell)}_{i\mu}}
\end{equation}
appears from chain-ruling through the layers, $x$-subscripts standing in for generic inputs. One can show $g, G \sim O_N(1)$ \cite{pehlevan2024lecture}, and with the feature kernels already $O_N(1)$, the condition
\begin{equation}
    \label{eqn:constraints_learning_predictions}
2a_1 + c = 0, \qquad 2a_\ell + c = 1 \,\,\, \forall \ell > 1
\end{equation}
ensures $\frac{d\phi(x)}{dt}\sim O_N(1)$: predictions move at finite speed, so learning happens in finite time.

\bigskip
\noindent \textbf{Features Evolve in Finite Time.}
Finally, the features themselves must move. For the first layer,
\begin{align}
 \frac{dz_{i\mu}^{(1)}}{dt}&=\frac{1}{N^{a_1}\sqrt{D}} \frac{dW_{ij}}{dt} x_{j\mu}
 = \frac{1}{N^{2a_1+c-d+1/2}}  \frac{\eta_0\gamma_0}{|\cD|}
 \sum_{\nu=1}^{|\cD|}\Delta_\nu g_{i\nu}^{(1)} \Phi_{\mu\nu}^{(0)},
\end{align}
all $N$-dependence sitting in the prefactor. Given \eqref{eqn:constraints_learning_predictions}, the requirement $dz^{(1)}/dt \sim O_N(1)$ forces
\begin{equation}
d=\frac12,
\end{equation}
and the same constraint propagates to every layer. Note in particular:
\CT{No feature learning}{if $d< \frac12$,}
which includes the NTK regime $d=0$.

\bigskip
\noindent \textbf{Summarizing the Constraints}. Assembling everything:
\begin{align}
    2a_1 + b_1 &= 0 \\
    2a_\ell + b_\ell &= 1 \quad \forall \ell > 1 \\
    2a_1 + c &= 0 \\
    2a_\ell + c &= 1 \quad \forall \ell > 1 \\
    d &= \frac12.
\end{align}
The solutions form a one-parameter family labeled by $a\in \bR$:
\begin{align}
(a_\ell, b_\ell, c_\ell) &= (a,1-2a,1-2a) \quad \forall \ell > 1 \\
(a_1, b_1, c_1) & = (a-\frac12, 1-2a, 1-2a).
\end{align}
Demanding additionally that $\eta \sim O_N(N^{2d-c})$ be $O_N(1)$ (no learning-rate retuning as you scale up) fixes $c=1$ and the solution is unique. This is the \emph{maximal update parameterization}, $\mu P$ \cite{yang2020feature}, and it is not just theory: it is how feature-learning behavior is preserved when scaling models in practice.

\section{Day 2: Neural Network Field Theory}
\label{sec:NNFT}

Day 1 kept bumping into field theory: correlators of network ensembles, generalized free fields at infinite width, $1/N$ interactions. The aim of those lectures was physics-for-ML. 

Today we reverse the logic, asking instead whether neural networks give a new approach to field theory.

\subsection{Essentials of NN-FT}

\subsubsection{What is a field theory?}

Let us start the day with an uncomfortably basic question:
\Q{What is a field theory?}
Some essential elements of a field theory include:
\begin{itemize}
    \item \textbf{Fields}: functions in some function space, or more generally distributions or sections of an appropriate bundle.
    \item \textbf{Correlation Functions} of those fields,
    \begin{equation}
        G^{(n)}(x_1,\dots,x_n) = \bk{\phi(x_1)\dots \phi(x_n)}.
    \end{equation}
\end{itemize}
You likely want to add to this list, but hold off for a moment. 

With just these ingredients,
\CT{Answer}{a FT is an ensemble of functions with a way to compute their correlators.}
We imagine doing so by a partition function
\begin{equation}
Z[J] = \bk{e^{\int d^dx J(x)\phi(x)}},
\end{equation}
agnostic, so far, about how $\bk{\cdot}$ is defined. The standard definition is Feynman's,
\begin{equation}
Z[J] = \int \cD \phi \, e^{-S[\phi] + \int d^dx J(x)\phi(x)},
\end{equation}
which requires an action $S[\phi]$ to define the density on functions. 
But an action is not what we specify when we specify a neural network. The NN data $(\phi_\theta, P(\theta))$ defines instead
\begin{equation}
\label{eqn:NNFT_Z}
Z[J] = \int d\theta\, P(\theta)\, e^{\int d^dx J(x)\phi_\theta(x)}.
\end{equation}
These are two different ways to define a field theory. 
When both exist for the same theory, we have dual descriptions of the statistics, as in the NNGP correspondence of Day 1. 

But crucially, \eqref{eqn:NNFT_Z} stands on its own: we can define, study, and compute with a field theory $(\phi_\theta, P(\theta))$ even if we do not know the action. Specifically, correlators can be computed from parameter space, and in some interacting theories they can be computed \emph{exactly} this way. Let me record the definition that carries the rest of the day.
\begin{definition}[Neural Network Field Theory]
An NN-FT is the data of an architecture $\phi_\theta:\bR^d\to\bR^D$ and a parameter density $P(\theta)$, with parameter-space partition function \eqref{eqn:NNFT_Z} and correlators.
\end{definition}
\noindent One may generalize to other types of fields by adding indices \cite{Maiti:2021fpy} or using Grassmann-valued neural networks \cite{Frank:2025zuk}, etc. The key point is that the theory is defined by an architecture and a density on its parameters.

\bigskip
Now, your objections. You wanted the definition of field theory to include $X$, for some
\begin{align}
X \in \{
    \text{Quantum} ,
    \text{ Lagrangian},
    \text{ Symmetries},
    \text{ Locality}, \dots
    \}.
\end{align}
The trouble is that for every such $X$ there is a thriving community of physicists who happily drop it: statistical field theories that are not Wick rotations of nice quantum theories, theories with no known Lagrangian, theories with no symmetries, non-local theories, etc. 

So I keep the minimal definition (fields and correlators), and everything else becomes an engineering problem:
\Q{Can I engineer my defining data to get FT + $X$?}
Whether your defining data is $S[\phi]$ or $(\phi_\theta,P(\theta))$, this is a natural question, and much NN-FT progress from the last few years involves answering it, $X$ by $X$; references for various $X$ are below.

\subsubsection{Free Theories by Design}

The first engineering target is the free theory, and Day 1 already did the work. Architectures of the form
\begin{equation}
\phi(x) = \frac{1}{\sqrt{N}}\sum_i w_i \varphi_i(x)
\end{equation}
with i.i.d. neurons become Gaussian as $N\to\infty$ by the CLT (generalized free fields), and the Euclidean-invariant input layer with its spectrum-shaping function $F$ lets us dial in the propagator we want. A key result was Scalar-net \eqref{eqn:scalar-net} with power spectrum:
\begin{equation}
G^{(2)}(p) = \frac{1}{p^2+m^2},
\end{equation}
the free scalar in $d$ dimensions, realized as an infinite-width network ensemble with uniform-in-the-momentum-ball hidden weights. 

The general recipe:
\CT{Recipe}{to engineer a free NN-FT, choose the architecture and density so the CLT applies, then shape $G^{(2)}$ to taste.}
Variants of this construction used for the free scalar (random Fourier features, in ML parlance) will reappear today as the worldsheet boson of the string and the spin waves of the XY model.

\subsubsection{Quantum Field Theory}

We have been resolutely Euclidean. In what sense can these theories be \emph{quantum}? In a QFT course one canonically quantizes and Wick rotates, and the quantum credentials are manifest. Here we have neither a Hilbert space nor an action, only Euclidean correlators, so we must ask
\Q{Given Euclidean correlators, can the theory be continued to a well-behaved Lorentzian quantum theory?}
This is the founding question of constructive field theory, and it has a famous answer \cite{Osterwalder:1973dx}.
\begin{theorem}[Osterwalder--Schrader Reconstruction]
A set of Euclidean correlators that are tempered distributions and obey Euclidean invariance, permutation symmetry, reflection positivity, and cluster decomposition analytically continues to a unitary Lorentzian quantum field theory satisfying the Wightman axioms.
\end{theorem}
\noindent Spelled out, the conditions are:
\begin{itemize}
\item \textbf{Euclidean Invariance} of the correlators, becoming Poincar\'e invariance after continuation.
\item \textbf{Permutation Invariance} of $G^{(n)}(x_1,\dots,x_n)$ in its arguments.
\item \textbf{Reflection Positivity}: choosing a Euclidean time $\tau$ and letting $R(x)$ reflect through $\tau=0$,
\begin{equation}
    G^{(2n)}(x_1,\dots,x_n,R(x_1),\dots,R(x_n)) \geq 0,
\end{equation}
the Euclidean shadow of unitarity and positive-norm states. (Necessary form; see \cite{Osterwalder:1973dx,Simmons-Duffin:2016gjk} for an elaboration.)
\item \textbf{Cluster Decomposition}: connected correlators vanish at infinite separation.
\end{itemize}
A pair $(\phi_\theta, P(\theta))$ whose correlators clear these hurdles defines a neural network \emph{quantum} field theory \cite{Halverson:2021aot}, in the sense that the O-S axioms determine a QFT. Permutation invariance is automatic in NN-FT; Euclidean invariance can be engineered by the absorption mechanism of Day 1; reflection positivity and cluster decomposition hold in examples \cite{Halverson:2021aot,Demirtas:2023fir} but systematic constructions beyond the case of quantum mechanics \cite{Ferko:2025ogz} remain an open problem.

\subsubsection{Symmetries}

Day 1's absorption mechanism \cite{Maiti:2021fpy} is a simple way to realize symmetries:
\begin{enumerate}
\item \textbf{Act.} Realize $g$-action on a network for arbitrary $g\in G$.
\item \textbf{Absorb.} Redefine parameters to absorb the $g$-action.
\item \textbf{Invariance.} Check that $d\theta\, P(\theta)$ does not transform.
\end{enumerate}  Cos-net realized the Euclidean group this way. Two upgrades arrive in the Recent Results below: conformal and Virasoro symmetry can be engineered \cite{Halverson:2024axc, Robinson:2025ybg, Capuozzo:2025ozt}, and the absorption mechanism has been promoted to a full theory of Ward identities, Schwinger-Dyson equations, and anomalies in parameter space \cite{Ferko:2026kkm}. Fermions and supersymmetry also now exist in NN-FT \cite{Frank:2025zuk}, via Grassmann-valued parameters, which today will give us ghosts.

\subsubsection{Interactions}

Free theories come from the CLT; interactions come from breaking it. Day 1's first lever was finite $N$, giving connected correlators at $O(1/N)$. The second lever, independence breaking, can also be made explicit. Suppose $(\phi_\theta, P(\theta))$ is an NNGP with Gaussian partition function
\begin{equation}
Z_{G}[J] = \int d\theta \,P(\theta)\, e^{\int d^dx J(x)\phi_\theta(x)}.
\end{equation}
Insert the operator associated to any local potential $V(\phi)$:
\begin{align}
Z[J] &= \int d\theta \,P(\theta)\, e^{ -\int d^dx V(\phi_\theta(x))} e^{\int d^dx J(x)\phi_\theta(x)} \\
&=: \int d\theta \,\tilde P(\theta) \,e^{\int d^dx J(x)\phi_{\theta}(x)}.
\end{align}
The architecture equation lets us trade the abstract insertion for a concrete function of parameters, defining a deformed density
\begin{equation}
\tilde P(\theta) = P(\theta)\, e^{ -\int d^dx V(\phi_\theta(x))}
\end{equation} 
The point is that since $\phi_\theta$ is given by an explicit architecture, inserting it yields a concrete function of the parameters that mixes them.
In general the deformation breaks the statistical independence utilized in the CLT, yielding interactions.

For $\phi^4$ theory: insert
\begin{equation}
e^{-\frac{\lambda}{4!}\int d^dx\, \phi_\theta(x)^4}
\end{equation}
into the scalar-net partition function (an IR regulator is needed for that particular architecture), deforming \eqref{eqn:scalar-net-densities} into a correlated density. See \cite{Demirtas:2023fir} for the full treatment. 

In summary, we have:
\CT{Interaction Mechanism}{from violating CLT assumptions, via $1/N$-corrections or independence breaking.}
Incidentally, this mechanism (deform the Gaussian density, keep the architecture) is how Liouville theory is constructed \cite{david2016liouville} in the probabilistic QFT community, a subject of study in the Simons Collaboration on Probabilistic Paths to QFT \cite{probabilistic_qft_2026}.

\subsection{Recent Results}
\label{sec:recent}

There are a number of recent results, especially since November 2025, utilizing NN-FT in the for-physics direction. These include conformal symmetry \cite{Halverson:2024axc}, defect CFTs \cite{Capuozzo:2025ozt}, Virasoro symmetry \cite{Robinson:2025ybg}, fermions and SUSY \cite{Frank:2025zuk, Jiang:2025lzm}, renormalization-group flow \cite{Howard:2024kfd}, transformer architectures \cite{Ageev:2026qyh}, an optimality result \cite{Zhang:2026tss} for different theories realizing the same GP limit, and the results of this section: universality \cite{Ferko:2026axm}, Liouville theory \cite{Ferko:2026axm}, the bosonic string \cite{Frank:2026bui}, topological sectors and the Kosterlitz-Thouless transition \cite{Ferko:2026ken}, and anomalies \cite{Ferko:2026kkm}. 

\subsubsection{Universality}
\label{sec:universality}

In Day 1 we discussed expressivity of neural networks, namely, whether they can represent or approximate functions of a given type. The answer was affirmative, established, e.g., via Cybenko's Universal Approximation Theorem.

A natural extension of this expressivity question to the context of NN-FT is 
\Q{Which field theories admit a NN-FT description?}
Answering the question requires being sufficiently precise about what we mean by a field theory. In the case of scalar field theories described using the machinery of constructive QFT, the answer is that NN-FT is universal \cite{Ferko:2026axm}, as we make precise in this section. 

First, what do we mean by a field theory? Known field theories exhibit fields that are ``rougher" than a function. To see this, recall that the kinetic term $\partial_\mu \phi \partial^\mu \phi$ is given by 
\begin{equation}
a^d\left(\frac{\phi(x + an)-\phi(x)}{a}\right)^2
\end{equation}
in a lattice theory with spacing $a$, and $n$ a lattice vector. 
Scaling to the continuum limit requires 
\begin{equation}
\phi(x+an)-\phi(x)\sim a^{1-d/2},
\end{equation} so that the kinetic term remains finite as $a\to 0$.
In $d=1$ this is Brownian roughness (continuous, nowhere differentiable), and in $d\geq 2$ typical configurations are not functions at all but \emph{tempered Schwartz distributions}, elements of $\cS'(\bR^d)$. Constructive field theory accordingly defines a Euclidean QFT as a probability distribution over $\cS'(\bR^d)$, with the OS axioms layered on top when one wants a continuation to a unitary theory in Lorentzian signature. 

Refining our question regarding the universality of NN-FT, we ask:
\Q{Do all probability distributions over $\cS'(\bR^d)$ admit a NN-FT description in terms of data $(\phi_\theta, P(\theta))$?}
To address the question, we should be precise about what a ``neural network description'' means. The cleanest picture is two probability spaces and one map between them, lifted from \cite{Ferko:2026axm}:
\begin{center}
\begin{tikzpicture}[>={Stealth[length=2.4mm]}, line width=0.8pt]
  \node (nn) at (0,0)   {$\big(\bR^{\bN},\,\mathcal{B}(\bR)^{\otimes\bN},\,P\big)$};
  \node (sp) at (6.5,0) {$\big(\cS'(\bR^d),\,\mathcal{B}(\cS'),\,P_\phi\big)$};
  \draw[->] (nn) -- node[above]{$\phi_\theta$} node[below]{arch.} (sp);
  \node[font=\footnotesize] at (0,-0.66)   {the NN parameters};
  \node[font=\footnotesize] at (6.5,-0.66) {the field theory};
\end{tikzpicture}
\end{center}
The two ends are both \textbf{probability spaces}, and the arrow between them is the architecture:
\begin{itemize}
\item the \textbf{field theory} (right) is a probability distribution $P_\phi$ over field configurations in $\cS'(\bR^d)$;
\item the \textbf{neural network parameters} (left) is a probability distribution $P$ over parameters $\theta\in\bR^{\bN}$.
\item the \textbf{architecture} $\phi_\theta$ takes parameters to a field configuration, pushing forward the parameter density as  $(\phi_\theta)_{*}P=P_\phi$.
\end{itemize}
In this perspective, a NN-FT is a field theory obtained by pushing a parameter density through an architecture, and the universality question is whether every field theory can be realized this way.

It can, for a clean reason. Both ends are standard Borel spaces, and the Borel isomorphism theorem guarantees the existence of an isomorphism between any two such spaces. That is, there exists $\phi_\theta$ such that 
\begin{equation}
\phi_\theta:\bR^{\bN}\;\xrightarrow{\ \sim\ }\;\cS'(\bR^d)
\end{equation}
is an isomorphism, so that given a $P_\phi$ there is a $P$ such that $(\phi_\theta)_{*}P=P_\phi$. One can take the architecture to be the Borel isomorphism itself.
Hence \cite{Ferko:2026axm}:
\begin{theorem}[Universality of NN-FT]
Any probability distribution over tempered distributions on $\bR^d$ (in particular any Euclidean QFT in the constructive sense, whether or not the OS axioms hold) admits a NN description $(\phi_\theta, P(\theta))$ with countably many parameters.
\end{theorem}
\noindent We emphasize that the architecture guaranteed by the Borel isomorphism, unlike many in machine learning, is one-to-one between parameters and functions. Such is the case when one takes the neurons to be a basis $\varphi_i$ for the function space.

\subsubsection{Liouville Theory}
\label{sec:liouville}

Liouville theory is an interacting 2d CFT with action given by  
\begin{equation}
S=\int_\Sigma d^2x\,\sqrt{g}\left(\frac{1}{4\pi}\partial^\mu\phi\,\partial_\mu\phi+\frac{Q}{4\pi}R\,\phi+\mu\, e^{2b\phi}\right),
\end{equation}
with background charge $Q = b + 1/b$ and $c_L = 1+6Q^2$. The primaries are vertex operators $V_\alpha = \,:e^{2\alpha\phi}:$ with $\Delta_\alpha = \alpha(Q-\alpha)$, and conformal symmetry fixes the three-point function up to the structure constants,
\begin{equation}
\bk{V_{\alpha_1}(x_1)V_{\alpha_2}(x_2)V_{\alpha_3}(x_3)} = \frac{C(\alpha_1,\alpha_2,\alpha_3)}{|x_{12}|^{2\Delta_{123}}|x_{13}|^{2\Delta_{132}}|x_{23}|^{2\Delta_{231}}},
\end{equation}
$\Delta_{ijk} = \Delta_i + \Delta_j - \Delta_k$. The structure constants are the content of the DOZZ formula \cite{Dorn:1994xn,Zamolodchikov:1995aa}: with $\gamma(x)=\Gamma(x)/\Gamma(1-x)$, $s=(\alpha_1+\alpha_2+\alpha_3-Q)/b$, and Zamolodchikov's special function $\Upsilon_b$,
\begin{equation}
C=\Big(\pi\mu\,\gamma(b^2)\,b^{2-2b^2}\Big)^{-s}\,\frac{\Upsilon_b'(0)\,\prod_{i=1}^3\Upsilon_b(2\alpha_i)}{\Upsilon_b\!\big(\textstyle\sum_i\alpha_i-Q\big)\,\prod_{i=1}^3\Upsilon_b\!\big(\textstyle\sum_j\alpha_j-2\alpha_i\big)},
\end{equation}
where, for $0<\mathrm{Re}\,z<Q$,
\begin{equation}
\log\Upsilon_b(z)=\int_0^\infty\frac{dt}{t}\left[\Big(\tfrac{Q}{2}-z\Big)^2 e^{-t}-\frac{\sinh^2\!\big[\big(\tfrac{Q}{2}-z\big)\tfrac t2\big]}{\sinh\tfrac{bt}{2}\,\sinh\tfrac{t}{2b}}\right].
\end{equation}
This irrational-$c$ structure constant is our target.

Recent work \cite{david2016liouville} (see \cite{chatterjee2024liouville} for an introduction) in the probabilistic QFT community established that Liouville theory is a rigorously defined interacting theory, and it turns out that the construction is effectively an NN-FT.    The architecture is a zero mode plus real spherical harmonics,
\begin{equation}
\phi_\theta = c+\sum_{\ell=1}^{L}\sum_{m=-\ell}^{\ell}a_{\ell,m}\,Y_{\ell,m},
\end{equation}
parameters $\theta = \{c, a_{\ell,m}\}$. The free field on the sphere obtained from Gaussian parameters drawn as 
\begin{equation}
a_{\ell,m}\sim\cN\Big(0,\ \frac{4\pi}{\ell(\ell+1)}\Big).
\end{equation} Interactions enter as advertised: deform the parameter density. 

Remarkably, in NN-FT all of the Liouville interaction can be loaded onto the \emph{zero mode}. Specifically, both interaction terms reduce to functions of the constant $c$, which we justify before stating the clean NN-FT answer. On the round sphere $R$ is constant, so the curvature coupling integrates against the fluctuations to zero ($\int_{S^2}\sqrt{g}\,R\,Y_{\ell,m}=0$ for $\ell\geq1$), leaving $\int_{S^2}\sqrt{g}\,\tfrac{Q}{4\pi}R\,\phi = 2Qc$, pure zero mode. The exponential, meanwhile, factorizes, $e^{2b\phi}=e^{2bc}\,e^{2bX}$, so its only zero-mode dependence is the scalar prefactor $e^{2bc}$ multiplying the fluctuation piece $M_X = \mu\int_{S^2}\sqrt{g}\,{:}e^{2bX}{:}$. Both terms therefore live entirely in $c$ (alongside the vertex factors $e^{2\alpha_i c}$), so only the zero mode's density gets deformed. Writing $t = e^{2bc}$, the density factorizes as 
\begin{equation}P(\theta) = P(t \mid a_{\ell,m})\, P(a_{\ell,m})\end{equation} 
where the NN parameters are $\theta = \{t, a_{\ell,m}\}$ and the conditional density of the zero mode is
\begin{equation}
P(t\mid a_{\ell,m}) \propto t^{s-1}\exp\big(-M_X(a_{\ell,m})\,t\big), \qquad s=\frac{\sum_i\alpha_i-Q}{b},
\end{equation}
where
\begin{equation}
M_X = \mu \int_{S^2} d^2x \sqrt{g}\, :e^{2b \sum a_{\ell,m} Y_{\ell,m}}: \,
\end{equation}
is the so-called Gaussian multiplicative chaos. The non-zero modes stay independent Gaussians, and every bit of statistical dependence, every bit of interaction, lives in how the zero mode is conditioned on them. The exponent $s$ collects the zero mode's three appearances: the curvature coupling (which the non-zero modes never see, since $\int \sqrt{g}\, Y_{\ell,m} R = 0$ on the round sphere), the vertex operator insertions, and the exponential interaction.

\begin{figure}[!ht]
\centering
\includegraphics[width=\columnwidth]{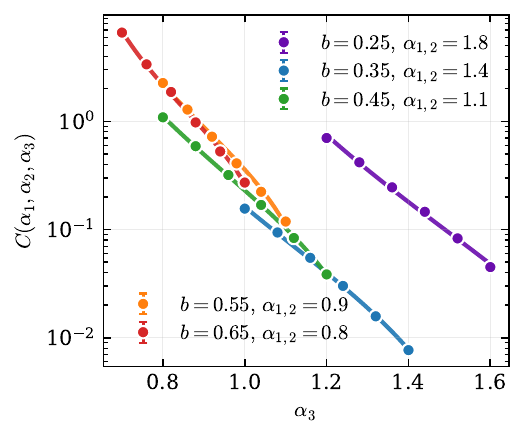}
\caption{The Liouville structure constant $C(\alpha_1,\alpha_2,\alpha_3)$ from NN correlators (points; $\alpha_1=\alpha_2$ fixed and $\alpha_3$ varied, for several $b$) versus the exact DOZZ formula (solid curves), which lies within all error bars. NN-FT run: $L=30$ spherical harmonics, $10$ experiments of $50{,}000$ runs, sphere pixelated at $100\times200$ points; the error bars (variance across experiments) are smaller than the markers \cite{Ferko:2026axm}.}
\label{fig:dozz}
\end{figure}

The computation \cite{Ferko:2026axm} is then honest Monte Carlo in parameter space: sample the $a_{\ell,m}$, normal-order via mollified ensemble variances, pixelate the sphere to evaluate $M_X$, sample $t$ from its conditional density, average.
A non-perturbative numerical computation at $L=30$ yields agreement with the DOZZ formula at the few-percent level across a range of $b$ and $\alpha_3$ (Fig.~\ref{fig:dozz}), with error bars from independent experiments smaller than the plot markers, degrading (as it should) near the Seiberg bound $\alpha \leq Q/2$ where DOZZ develops zeros.

\subsubsection{The Bosonic String}
\label{sec:string}

We are at a strings school, so we should ask the question:
\Q{Is string theory an NN-FT?}
For the bosonic string, the answer is yes \cite{Frank:2026bui}: the gauge-fixed worldsheet theory is realized by ensembles of infinite-width networks and allows for computation of the Virasoro-Shapiro and Veneziano amplitudes. (Worldsheet Virasoro symmetry in NN-FT is developed in the complementary \cite{Robinson:2025ybg}.)

\medskip
\noindent\textbf{Free Boson.} The embedding fields $X^\mu(z,\bar z)$, $\mu = 1,\dots,D$, on the plane $z = x_1 + i x_2$ are given the architecture
\begin{equation}
\label{eqn:string_arch}
X^{\mu}(z)=\frac{C}{\sqrt{N}}\sum_{i=1}^{N}\frac{a_{i}^{\mu}}{|W_{i}|}\cos\Big(\tfrac{1}{2}(W_{i}z+\bar{W}_{i}\bar{z})+c_{i}\Big),
\end{equation}
with i.i.d. draws
\begin{equation}
a_i^\mu \sim \cN(0,\sigma_a^2), \quad c_i \sim \text{Unif}[-\pi,\pi], \quad W_i \sim \text{Unif}\big(B^2_\Lambda\setminus B^2_\epsilon\big),
\end{equation}
the complex hidden weights $W_i$ uniform on an annulus in momentum space: $\Lambda$ a UV cutoff, $\epsilon$ an IR cutoff. This is the Euclidean-invariant layer of Day 1 cast into the context of a $2d$ worldsheet, with spectrum-shaping factor $1/|W_i|$. Averaging over parameters, the two-point function is a Bessel integral that lands on
\begin{equation}
\bk{X^\mu(z) X^\nu(w)} = \alpha'\delta^{\mu\nu}\Big[-\log r+\log\frac{2\Lambda}{\epsilon\, e^{\gamma}}+\cdots\Big], \quad r = |z-w|,
\end{equation}
the free-boson logarithm for $1/\Lambda \ll r \ll 1/\epsilon$, i.e. $G^{(2)}(p) = 2\pi\alpha'\,\delta^{\mu\nu}/p^2$ on the annulus. By the CLT, the $N\to\infty$ ensemble is the Gaussian theory with exactly this propagator: the gauge-fixed Polyakov boson, \emph{defined} by its statistics, no worldsheet action ever written down. Worldsheet Euclidean invariance comes from the $(W, c)$ densities, spacetime rotation invariance from the Gaussian on $a^\mu$. A zero mode $X_0^\mu \sim \cN(0,\sigma_0^2)$ may be appended.

Setting the normalization $C=1$ and matching coefficients yields the relation:
\begin{equation}
\alpha'=\frac{\sigma_a^2}{\Lambda^2-\epsilon^2}.
\end{equation}
The annulus radii are regulators; the only physical dial in the construction is $\sigma_a$, and the only physical parameter of the bosonic string is $\alpha'$. Other architectures may also realise the free boson, with output weight variance playing a similar role.

\medskip
\noindent\textbf{Ghosts.} The $bc$ system requires anticommuting fields, which NN-FT now supplies \cite{Frank:2025zuk}: take the same random-feature architecture but with Grassmann-valued output weights $\beta_k, \chi_k$ and a Grassmann Gaussian density $P \propto \exp(\beta_k\chi_k/\sigma^2)$, so Berezin integration gives $\bk{\beta_k \chi_l} = \sigma^2\delta_{kl}$. With phases arranged appropriately, the parameter-space average produces
\begin{equation}
\bk{b(z)\, c(w)} = \frac{1}{z-w}+\cdots,
\end{equation}
the standard ghost propagator, with corrections vanishing at large $\Lambda r$; see \cite{Frank:2026bui} for details. Ensembles of networks with Grassmann parameters: the worldsheet ghost system.

\medskip
\noindent\textbf{Amplitudes in parameter space.} Having realized the free boson and the $bc$-ghosts, we have realized the essential worldsheet fields of the gauge-fixed bosonic string and should be able to compute amplitudes.

The four-tachyon closed-string amplitude is the gauge-fixed correlator on the sphere 
\begin{equation}
\cA^{(4)}=\frac{1}{g_s^2}\,\frac{1}{\text{Vol}(SL(2,\bC))}\Big\langle \prod_{i=1}^4 V(p_i)\Big\rangle, \quad V(p)=g_s \cZ(p)\!\int\! d^2z\, e^{ip\cdot X(z)},
\end{equation}
where the expectation is the \emph{parameter-space} expectation. Written out, the amplitude is an honest finite-dimensional integral over the network parameters,
\begin{equation}
\begin{split}
\cA^{(4)}={}&\frac{g_s^2}{\text{Vol}(SL(2,\bC))}\int [d\mu_X]\Big(\prod_{\alpha=1}^4 d^2z_\alpha\Big)\\
&\times\,\cZ(p_1)\cdots\cZ(p_4)\,\exp\!\Big(i\sum_{j=1}^4\sum_{\mu=1}^D p_j^\mu X^\mu(z_j)\Big),
\end{split}
\end{equation}
taken against the explicit per-neuron measure
\begin{equation}
[d\mu_X]=\prod_{i=1}^N\bigg[\frac{d^2W_i}{\text{Vol}(B^2_\Lambda\setminus B^2_\epsilon)}\,\frac{dc_i}{2\pi}\,\prod_{\mu=1}^D\frac{da_i^\mu}{\sqrt{2\pi\sigma_a^2}}\,e^{-(a_i^\mu)^2/2\sigma_a^2}\bigg].
\end{equation}
No path integral and, to emphasize a key point, no Monte Carlo either: the computation is analytic. The Gaussian integral over the output weights $a_i^\mu$ is exact; the remaining $W, c$ integrals factorize over i.i.d. neurons into $[\Phi_N]^N \to \exp(N\log \Phi_N)$, which exponentiates at large $N$ into single-neuron expectation values. Evaluating them, the cross terms assemble precisely into the Koba-Nielsen factor
\begin{equation}
\prod_{r<s}|z_r-z_s|^{\alpha'\, p_r\cdot p_s},
\end{equation}
while the self-contractions (we never normal-ordered) are absorbed into the renormalization $\cZ(p)$. The zero-mode Gaussian contributes $e^{-\sigma_{\text{eff}}^2(\sum_r p_r)^2/2}$, which as $\sigma_{\text{eff}}\to\infty$ becomes $\delta^D(\sum_r p_r)$: momentum conservation emerges from translation invariance of the network density, in the decompactification limit of the zero mode. The result is the paper's final expression: the standard Virasoro-Shapiro amplitude, now obtained as a correlator in network parameter space,
\begin{equation}
\boxed{\ \cA^{(4)}=\frac{g_s^2\,\delta^D\!\big(\sum_r p_r\big)}{\text{Vol}(SL(2,\bC))}\int\Big(\prod_{\alpha=1}^4 d^2z_\alpha\Big)\prod_{r<s}|z_r-z_s|^{\alpha'\, p_r\cdot p_s}\ }
\end{equation}
with its crossing symmetry, Regge behavior, and famously soft UV.

For open strings, put the worldsheet on the upper half plane and enforce the Neumann condition by the image trick \emph{at the level of the architecture},
\begin{equation}
X_N^\mu(z,\bar z) = \tfrac{1}{\sqrt{2}}\left[X^\mu(z,\bar z) + X^\mu(\bar z, z)\right],
\end{equation}
which is symmetric under $z\leftrightarrow\bar z$, i.e. $X_N^\mu(z,\bar z)=X_N^\mu(\bar z,z)$. Differentiating then gives $\partial X_N^\mu=\bar\partial X_N^\mu$ on the boundary $z=\bar z$, which is the Neumann condition $(\partial-\bar\partial)X_N^\mu|_{z=\bar z}=0$. The image term simultaneously doubles the boundary propagator. Similar steps may be followed to compute the Veneziano amplitude. 

The construction is also architecture-robust: any $X^\mu = \frac{1}{\sqrt{N}}\sum_i a_i^\mu \varphi_i(z)$ whose neurons have a $1/p^2$ spectrum gives the same GP, with $\sigma_a$ always playing the role of $\alpha'$. Different architectures giving the same GP limit may differ at finite $N$, where $1/N$ non-Gaussianities would correct the amplitudes.

\subsubsection{Topological Defects and the BKT Transition}
\label{sec:BKT}

Everything so far has been built from continuous parameters. A natural discrete question is:
\Q{What about topological defects?}
A Gaussian random-feature field is a smooth, single-valued, real-valued object. If the target is compact ($\phi \sim \phi + 2\pi r$), then winding modes and vortices are simply not in the ensemble, no matter how cleverly you shape the spectrum. This mirrors the path integral, where topological sectors enter through the choice of field space and measure, not through Gaussian fluctuations. The resolution of \cite{Ferko:2026ken} is correspondingly structural: enlarge the latent space to include \emph{discrete} parameters $Q$ labeling the topological quantum numbers,
\begin{equation}
\bk{\cO} = \sum_Q \int d\theta\, P(\theta, Q)\, \cO[\phi_{\theta, Q}],
\end{equation}
which gives us a 
\CT{Mechanism}{the sum over topological sectors is a sum over discrete latent variables of the network ensemble.}
Discrete latents are completely standard in ML, e.g. Gaussian mixture models or mixture of experts utilized in LLMs.

One flagship application is a famous example of a topological phase transition, the Berezinskii-Kosterlitz-Thouless transition of the 2d compact boson / XY model, driven by vortex unbinding. Concretely, it is a model of a periodic field $\theta(x)\sim\theta(x)+2\pi$ with action
\begin{equation}
S[\theta]=\frac{K_0}{2}\int d^2x\,(\nabla\theta)^2,\qquad K_0=\frac{\rho_{s,0}}{T},
\end{equation}
where $T$ is the temperature, $\rho_{s,0}$ the bare spin stiffness, and $K_0$ the dimensionless stiffness controlling the algebraic decay ($\eta=\frac{1}{2\pi K_0}$). Compactness admits configurations of nonzero winding, $\oint_\gamma \nabla\theta\cdot d\boldsymbol\ell = 2\pi m$ with $m\in\bZ$ (the vortices), which no smooth Gaussian field can reach; these are precisely the discrete sector $Q$ above. We split the architecture as 
\begin{equation}
\theta(x) = b\,\theta_{\text{sw}}(x) + \theta_v(x),
\end{equation}
where $\theta_\text{sw}$ is the topologically trivial, smooth part of the XY spin angle, $\theta_v$ the singular winding part, and $b$ is related to the temperature.
We put the theory on an $L\times L$ torus and take the spin waves to be random Fourier features
\begin{equation}
\theta_{\text{sw}}(x)=\frac{A}{\sqrt{N}}\sum_{j=1}^{N}\cos(k_j\cdot x+\gamma_j),
\end{equation}
with $k_j = 2\pi n_j/L$ and $n_j$ drawn $\propto 1/|n_j|^2$. The vortices are the discrete sector, characterized by positions and charges $\{(x_a, m_a)\}$, with $m_a = \pm 1$ and total vortex number zero $\sum_a m_a = 0$, entering the architecture through the singular harmonic function
\begin{equation}
\theta_v(x)=\sum_{a=1}^{N_v}m_a\arg(x-x_a),
\end{equation}
with the discrete-sector density a neutral Coulomb gas,
\begin{equation}
P_{\text{vort}}\propto y^{N_v}\exp\Big[2\pi K_0\sum_{a<b}m_a m_b\log (r_{ab}+a_c)\Big],
\end{equation}
$y$ the vortex fugacity ($y\sim e^{-\beta E_c}$, the core-energy Boltzmann weight), $a_c$ a core regulator, $K_0$ the stiffness tied to $b$. The two sectors are statistically independent. Each sector may be sampled numerically: simple exact sampling for the spin wave, a Markov chain (displace a vortex, insert or delete a neutral pair) for the vortices.

\medskip
What BKT phenomena can we compute in this NN-FT?

\begin{figure}[!ht]
\centering
\includegraphics[width=\columnwidth]{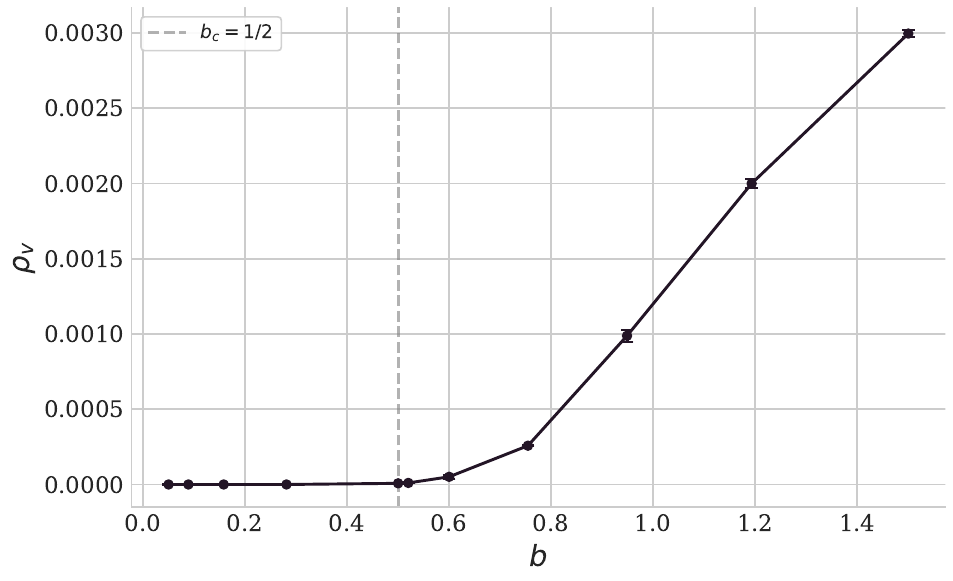}
\caption{Vortex density $\rho_v$ versus the coupling $b$ (the dashed line marks $b_c=\tfrac12$): consistent with zero below the transition, rising steeply above it as vortex--antivortex pairs unbind. From \cite{Ferko:2026ken}.}
\label{fig:vortex_density}
\end{figure}

At $N\to\infty$ the spin-spin correlator of the spin-wave sector may be computed directly in parameter space, giving
\begin{equation}
\bk{e^{ib\theta_{\text{sw}}(x)}e^{-ib\theta_{\text{sw}}(0)}} \sim |x|^{-\eta}, \qquad \eta = b^2\,.
\end{equation}
The BKT dictionary fixes temperature and stiffness in terms of the coupling, and the universal criticality condition $\eta_c=1/4$ locates the transition (units $\rho_{s,0}=1$):
\begin{equation}
K_0=\frac{1}{2\pi b^2},\quad T=2\pi b^2,\quad \eta_c=\tfrac14 \ \Longrightarrow\ \boxed{\,b_c=\tfrac12,\ \ T_c=\tfrac{\pi}{2}\,}.
\end{equation}
The numerics then check the full BKT phenomenology:
\begin{itemize}
\item \textbf{Below $T_c$}: fitted $\eta$ matches $b^2$ to better than a percent across the critical line, hitting the universal $\eta = 1/4$ at $b_c$ (Table~\ref{tab:bkt_eta}).
\item \textbf{Spin waves alone never transition}: Without vortices, the power law two-point function persists at all $b$, no transition.
\item \textbf{Above $T_c$}: the full two-point function
 decays exponentially,
\begin{equation}
G_2(r) = \bk{e^{i\theta(x)}e^{-i\theta(0)}} \propto e^{-r/\xi}%
\end{equation}
with the correlation length exhibiting the essential singularity
\begin{equation}
\xi(b)\sim \exp\!\Big(\frac{c}{\sqrt{b^2 - b_c^2}}\Big)
\end{equation}
of the BKT transition.
\item \textbf{Vortex proliferation}: the measured vortex density rises steeply above $b_c$ (Fig.~\ref{fig:vortex_density}); real-space temperature-dependent vortex maps  are given in \cite{Ferko:2026ken}
\item \textbf{Universal stiffness jump}: the renormalized stiffness $\Upsilon_R$ (the helicity modulus), extracted spectrally from the full field's small-momentum modes, tracks $K_0$ below the transition and collapses above it, crossing the Nelson-Kosterlitz line
\begin{equation}
\Upsilon_R(T_c^-)=\frac{2}{\pi}\,T_c
\end{equation}
at the transition --- the universal jump.
\end{itemize}
\begin{table}[!ht]
\centering
\begin{tabular}{ccc}
\hline
$b$ & $\eta_{\text{exp}}=b^2$ & $\eta_{\text{fit}}$ \\
\hline
0.050 & 0.00250 & 0.00252 \\
0.089 & 0.00791 & 0.00796 \\
0.158 & 0.02500 & 0.02517 \\
0.281 & 0.07906 & 0.07963 \\
0.500 & 0.25000 & 0.25204 \\
\hline
\end{tabular}
\caption{Below-$T_c$ power-law exponents: the fitted exponent $\eta_{\text{fit}}$ agrees with the analytic prediction $\eta_{\text{exp}}=b^2$ to better than $1\%$ across the critical line. From \cite{Ferko:2026ken}.}
\label{tab:bkt_eta}
\end{table}
These are the main phenomena of the BKT transition, realized explicitly in the context of a NN-FT with vortices.

\subsubsection{Anomalies and Ward Identities}
\label{sec:anomalies}

The last result is fundamental to NN-FT, and answers a question hovering since Day 1:
\Q{How does one understand symmetries, anomalies, and conservation laws in NN-FT?}
In the path integral, Ward identities and anomalies follow from a change of variables: act with a symmetry on the field, ask whether the integral and its measure are invariant, and read off a conservation law when they are, or an anomaly when the measure is not. 

The result of \cite{Ferko:2026kkm} is that NN-FT has its own integral, over parameters, and the same change-of-variables logic applied \emph{there} generates analogous machinery of Ward identities and anomalies (with Schwinger-Dyson (SD) equations as the more general case where the transformation is not necessarily a symmetry). The formalism may be applied even at finite width.

\medskip
\noindent\textbf{The Master Identity.} Let $p(\theta)>0$ be the parameter density and define the \textbf{score}
\begin{equation}
s_a(\theta) = \partial_a \log p(\theta),
\end{equation}
a centerpiece of modern generative modeling (the score-based generative models of \cite{song2019generative, song2020score} learn precisely this object, building on the diffusion framework of \cite{sohl2015deep}), here drafted into field theory. 

The derivation is a one-liner. For any smooth observable $\cO(\theta)$, the integral of a total derivative over parameter space vanishes up to boundary flux; expanding with the product rule, $\partial_a(p\,\cO) = p\,(s_a\,\cO + \partial_a\cO)$ gives one identity per parameter,
\begin{equation}
\label{eqn:SD_scalar}
\bk{\partial_a \cO} = -\bk{s_a(\theta)\,\cO},
\end{equation}
boundary terms dropped. This result is exact, with no continuum limit, locality, or Gaussianity assumed. Contracting against an arbitrary smooth vector field $\xi = \xi^a\partial_a$ on parameter space (send $\cO\to\xi^a\cO$, apply the product rule once more, and sum on $a$) promotes this to the master identity
\begin{equation}
\label{eqn:SD_master}
\Big\langle \sum_a \xi^a \partial_a \cO \Big\rangle = -\bk{B(\theta)\,\cO} + \text{(boundary)},
\end{equation}
where
\begin{equation}
\label{eqn:breaking_function}
B(\theta) = \underbrace{\sum_a \xi^a s_a}_{\text{score}} \;+\; \underbrace{\sum_a \partial_a \xi^a}_{\text{Jacobian}}
\end{equation}
is the \textbf{breaking function} of the flow $\xi$. No symmetry has been assumed: every flow has one. Field variations are induced by parameter flows, $\delta_\xi\phi(x) = \sum_a \xi^a\partial_a\phi(x)$, and taking $\cO$ to be a product of fields turns \eqref{eqn:SD_master} into $n$-point SD equations, with the score insertion playing the role that the equation-of-motion insertion plays in field space.

\medskip
\noindent\textbf{Ward Identities and Conservation Laws.} A parameter fluctuation varies the generating functional by
\begin{equation}
\label{eqn:deltaZ}
\begin{aligned}
\delta_\xi Z[J] &= \int d\theta\, p(\theta)\,\Big(\int d^dx\, J(x)\,\delta_\xi\phi(x)\Big)\, e^{\int J\phi} \\
&= -\bk{B(\theta)\, e^{\int d^dx\, J\phi}}\, Z[0] + (\text{boundary}),
\end{aligned}
\end{equation}
the second equality being the master identity \eqref{eqn:SD_master} with $\cO = e^{\int J\phi}$. If the flow is a symmetry of the ensemble, then
\begin{equation}
\delta_\xi Z[J] = 0
\end{equation} and \eqref{eqn:SD_master} becomes a Ward identity. The case of unbroken symmetry has a crisp characterization: since $p\,B = \sum_a \partial_a(\xi^a p)$,
\CT{Conservation}{$B = 0 \iff \partial_a\, j^a = 0$, where $j^a = \xi^a p$ is a conserved current on \emph{parameter space}.}
The current is not a local current on spacetime: it is a divergence-free vector field on $\bR^{|\theta|}$. 

When the theory \emph{does} have a local Lagrangian description, the familiar spacetime current is recovered by pushing forward: averaging $B(\theta)$ over the fiber of parameters mapping to a fixed field configuration,
\begin{equation}
B_{\text{eff}}[\phi] = \frac{\int d\theta\, p(\theta)\, B(\theta)\, \delta[\phi - \phi_\theta]}{\int d\theta\, p(\theta)\, \delta[\phi-\phi_\theta]},
\end{equation}
and locality then converts $B_{\text{eff}}$ into $\int d^dx\, \partial_\mu j^\mu_{\text{eff}}$. The chain
\begin{equation}
j^a \xrightarrow{\ \text{div}\ } p\,B(\theta) \xrightarrow{\ \text{fiber avg}\ } B_{\text{eff}}[\phi] \xrightarrow{\ \text{locality}\ } \partial_\mu j^\mu_{\text{eff}}(x)
\end{equation}
holds at any width and architecture for the first two arrows; only the last needs locality. So the formalism is rich enough to handle even non-local NN-FTs, where no $j^\mu(x)$ exists to write down.

The decomposition \eqref{eqn:breaking_function} is itself familiar by analogy to the path integral. The score term vanishes when the density is invariant along the flow (explicit breaking lives there), while the Jacobian term is density-independent. It measures the failure of the flat measure $d\theta$ to be invariant, a finite-dimensional avatar of the Fujikawa anomaly. 

Let us work out an ML sanity check. The input enters the network only through the first hidden layer's pre-activation
\begin{equation}
z_m(x) = \sum_j W^1_{mj}\,x_j + b^1_m,
\end{equation}
so ask whether the ensemble is invariant under rotations of the input, $x\mapsto (I+\epsilon\,\omega)x$ with $\omega$ antisymmetric. Such a rotation is undone by the compensating flow on the first-layer weights $\xi^{W^1_{mj}} = \sum_k W^1_{mk}\,\omega_{kj}$ (all other $\xi^a=0$), which leaves every $z_m(x)$, and hence every output, unchanged; we evaluate its breaking function. The Jacobian piece dies immediately, $\sum_a\partial_a\xi^a = n_1\,\mathrm{tr}\,\omega = 0$ by antisymmetry. The score piece, for a Gaussian i.i.d. density with $s_{W^1_{mj}} = -W^1_{mj}/\sigma^2$, is
\begin{equation}
B = \sum_{m,j,k} W^1_{mk}\,\omega_{kj}\,s_{W^1_{mj}} = -\frac{1}{\sigma^2}\sum_m \sum_{j,k} W^1_{mj}\,W^1_{mk}\,\omega_{kj} = 0,
\end{equation}
since the symmetric matrix $\sum_m W^1_{mj}W^1_{mk}$ is contracted against antisymmetric $\omega_{kj}$. Both pieces vanish: the Gaussian initialization is \emph{exactly} rotation invariant, and the master identity collapses to the unbroken rotational Ward identity. (Nothing hinges on the Gaussian specifically; any density depending on the weights only through the Gram matrix $G_{mn}=\sum_j W^1_{mj}W^1_{nj}$ does the job.) 

The same toolkit, pointed at the query/key rotation $W_Q\to RW_Q$, $W_K\to RW_K$ of the attention mechanism, instead returns a \emph{gauge} redundancy rather than a physical symmetry, yielding the trivial Ward identity such a redundancy deserves; we leave the details to \cite{Ferko:2026kkm}.

The scale anomaly of massless $4d$ $\phi^4$ similarly emerges from this formalism, with the one-loop beta function falling out of the parameter-space Ward identity,
\begin{equation}
\beta(\lambda)=\frac{3\lambda^2}{16\pi^2}+O(\lambda^3).
\end{equation}
Crucially, the appearance of the $\beta$-function arises from a Gaussian integration by parts in parameter space 
that converts the score insertion of the breaking function into the \emph{connected} four-point function, giving rise to the $\beta$-function via the usual one-loop diagram.

\medskip
\noindent\textbf{The Weyl anomaly and Critical Dimension.} Let us compute the critical dimension of the bosonic string. We work on the round $S^2$: it is the simplest closed (genus-zero) worldsheet and its Laplacian spectrum is the clean tower of spherical harmonics. The free boson is a harmonic random-feature field at $L\to \infty$,
\begin{equation}
X^\mu(\hat n) = \sum_{\ell=0}^{L}\sum_{m=-\ell}^{\ell} a^\mu_{\ell m}\, Y_{\ell m}(\hat n), \qquad \sigma_\ell^2 = \frac{2\pi\alpha'}{\ell(\ell+1)},
\end{equation}
while the $bc$ ghosts, holomorphic $b\equiv b_{zz}$ (spin-weight $+2$) and $c\equiv c^z$ (spin-weight $-1$), expand in spin-weighted spherical harmonics ${}_sY_{\ell m}$ \cite{NewmanPenrose1966,Goldberg1967}, which exist only for $\ell\geq|s|$,
\begin{equation}
b(\hat n) = \sum_{\ell=2}^{L}\sum_{m=-\ell}^{\ell}\beta_{\ell m}\,{}_{-2}Y_{\ell m}^*(\hat n), \qquad c(\hat n) = \sum_{\ell=1}^{L}\sum_{m=-\ell}^{\ell}\chi_{\ell m}\,{}_{-1}Y_{\ell m}(\hat n),
\end{equation}
with a Grassmann pair density on the paired ($\ell\geq2$) modes,
\begin{equation}
p_{bc} \propto \exp\!\Big(\frac{\mu_\ell}{4\pi}\,\beta_{\ell m}\chi_{\ell m}\Big), \qquad \mu_\ell = \sqrt{(\ell-1)(\ell+2)}\quad(\ell\geq 2).
\end{equation}
The constraint $\ell\geq|s|$ starts the $b$-sum at $\ell=2$ and the $c$-sum at $\ell=1$, so the three unpaired $\ell=1$ ghost modes are the conformal Killing vectors, saturated by the usual $c\bar c$ dressing. The antiholomorphic $\bar b,\bar c$ are an independent copy (spin-weights $-2,+1$), which is why the ghost Jacobian below carries a factor of two.

A constant Weyl rescaling $g\to e^{2\epsilon}g$ scales the Laplacian spectrum and hence maps this NN-FT to a NN-FT on a different background; absorbing the transformation into parameters gives the linear flows
\begin{equation}
\xi^{a^\mu_{\ell m}}=-a^\mu_{\ell m},\qquad \xi^{\beta_{\ell m}}=+\beta_{\ell m},\qquad \xi^{\chi_{\ell m}}=-2\chi_{\ell m}.
\end{equation}
In the breaking function, the score contributions cancel sector by sector against the Weyl shift of the spectrum --- the would-be explicit breaking eats itself --- leaving \emph{only} the Jacobian. And via direct computation, one realizes the Jacobian is a pure mode count:
\begin{equation}
B^{(X)}_{\text{Jac}}=-D\Big[\sum_{\ell=1}^L(2\ell+1)+1\Big],\qquad B^{(bc)}_{\text{Jac}}=2\sum_{\ell=2}^L(2\ell+1)+12.
\end{equation}
The counts diverge as $L\to\infty$; regularizing with zeta functions, $\zeta_X(0) = -2/3$ and $\zeta_{bc}(0) = -5/3$, the total is
\begin{equation}
B^{\text{reg}}_{\text{anom}} = -D\,[\zeta_X(0)+1] + 2\zeta_{bc}(0) + 12 = -\frac{D}{3}-\frac{10}{3}+12 = -\frac{1}{3}(D-26),
\end{equation}
and demanding Weyl invariance of the ensemble yields
\CT{Critical dimension}{$D=26$, from a count of network modes.}
The usual derivations get $26$ from a normal-ordering constant, or the Polyakov measure, or $c_{\text{matter}} + c_{\text{ghost}} = 0$. Here it is the regularized mismatch between how bosonic and Grassmann parameter measures respond to a flow on parameter space.

\section{Day 3: Applied AI for Strings}
\label{sec:day3}

Days 1 and 2 were theory, focused on understanding the essentials of neural networks and the relationship with field theory. 
Today we turn to practice. I want to give you working knowledge of some of the primary AI techniques used in string theory: enough ML to read the papers, plus applications of each.

I am deliberately leading with a modern topic, agentic reasoning models, because it changes how easily the other techniques can be implemented. My hope is not just that you hear these applied lectures, but also that you find them useful in your research. Agents grease the wheels for applying the other techniques.

ML techniques have been applied in string theory for almost a decade, beginning with \cite{He:2017aed,Krefl:2017yox,Ruehle:2017mzq,Carifio:2017bov} (see \cite{Ruehle:2020jrk,He:2018jtw,gukov2024rigor} for reviews). 
 The techniques we focus on are supervised learning, reinforcement learning, physics-informed networks, and conjecture generation. The field has had an annual meeting, \emph{String Data}, that is in its tenth edition in $2026$. Like \emph{Strings}, the physics is much broader than string theory, with a number of extensions into field theory and even pure math. A far from exhaustive\footnote{If I have missed your paper, please feel free to write, with BibTeX included.} sample of the literature, organized by type of learning used, is:
\emph{Supervised regression and classification} on string data: CICY and Kreuzer--Skarke datasets, line-bundle cohomology, string standard models, and holographic, toric-volume, and BPS-spectrum observables \cite{Bull:2018uow,Erbin:2020tks,Klaewer:2018sfl,Brodie:2019dfx,Deen:2020dlf,Hashimoto:2018ftp,Choi:2023rqg,Gukov:2024opc}. \emph{Reinforcement learning and genetic search} over flux vacua and heterotic backgrounds \cite{Halverson:2019tkf,Cole:2019enn,Krippendorf:2021uxu,Cole:2021nnt,Abel:2014xta,Constantin:2021for,Abel:2021rrj,Abel:2021ddu,Berglund:2023ztk,He:2026rdb,Harvey:2021oue}. \emph{Generative models} \cite{Cheng:2019xrt,Krippendorf:2020gny,Halverson:2020opj}. \emph{Physics-informed and variational solvers}, e.g. for Ricci-flat Calabi--Yau metrics \cite{Ashmore:2019wzb,Douglas:2020hpv,Anderson:2020hux,Jejjala:2020wcc,Larfors:2021pbb,Larfors:2022nep,Gerdes:2022nzr}. And, in low-dimensional topology, networks that learn knot invariants and yield rigorous results \cite{Jejjala:2019kio,Craven:2021ckk,Gukov:2020qaj,davies2021advancing,gukov2024rigor,Gukov_2025}.

\subsection{Agents for Everything}
\label{sec:agents}

In practice, much of the time in ML-for-strings research is spent on plumbing, building the dataset, wrangling it into tensors, writing the environment, tuning hyperparameters, debugging, regenerating plots, etc. Many techniques in the rest of today's lecture historically came with weeks of this. However, 
\CT{Key Point}{The plumbing can now be outsourced.}
Agents have significantly lowered the cost of trying things, and rapid prototyping of ML-for-physics ideas is possible in a way it simply was not even two years ago.
\Q{What exactly is an agent?}
One can mean a variety of things, but we will say:
\A{An ML algorithm that can carry out actions.}
We will eventually get to the traditional notion realized in a reinforcement learning algorithm, but the recent progress centers on \emph{agentic reasoning models}, e.g. \cite{openai2024o1,deepseek2025r1}: LLMs that can reason, plan, and act on an environment, through tools, in an \emph{agentic loop}.

\begin{figure}
\centering
\resizebox{\columnwidth}{!}{%
\begin{tikzpicture}[
  box/.style={draw, thick, rounded corners, align=center, minimum width=2.9cm, minimum height=0.95cm, font=\small},
  arr/.style={-{Stealth[length=2.5mm]}, thick}
]
\node[box] (ctx) at (0,0) {\textbf{Context}\\ \footnotesize task, history, memory};
\node[box] (llm) at (4.4,1.5) {\textbf{Reasoning model}\\ \footnotesize think, plan, decide};
\node[box] (act) at (8.8,0) {\textbf{Action}\\ \footnotesize run code, search, \\ \footnotesize compute, spawn agents};
\node[box] (env) at (4.4,-1.7) {\textbf{Environment}\\ \footnotesize shell, GPU, arXiv, files};
\draw[arr] (ctx) to[bend left=18] (llm);
\draw[arr] (llm) to[bend left=18] (act);
\draw[arr] (act) to[bend left=18] (env);
\draw[arr] (env) to[bend left=18] node[below left=-2pt]{\footnotesize observations} (ctx);
\draw[arr, dashed] (llm) -- ++(0,1.5) node[above, font=\footnotesize, align=center]{artifact: code, plots,\\ derivation, draft};
\draw[arr, dashed] (-2.4,1.3) node[left, font=\footnotesize]{goal} to[bend left=10] (ctx);
\end{tikzpicture}}
\caption{The agent loop. A reasoning model repeatedly thinks, acts on an environment through tools, and folds the resulting observations back into its context, terminating in an artifact. Orchestration patterns (subagents, critics, verification loops, etc.) are built from different copies of this loop.}
\label{fig:agent_loop}
\end{figure}
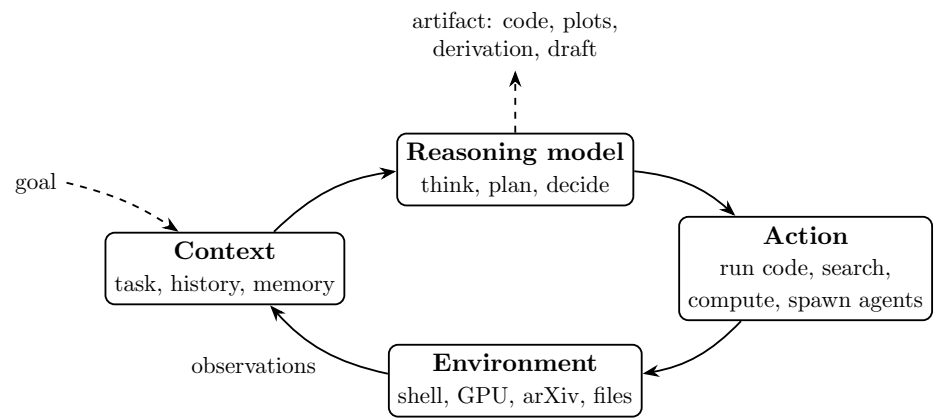

Such loops can have many forms, but a characteristic one is in (Fig.~\ref{fig:agent_loop}). There, the agent is a reasoning model placed in an environment. The model receives a context (the task, the conversation so far, observations), thinks, and then either answers or takes an \emph{action}: write and execute code, search the literature, call a computer algebra system, read a file, spawn a subagent to handle a subtask. The environment returns an observation (program output, an error message, a paper abstract), which is appended to the context, and the loop continues until the model judges the task complete and emits an artifact such as a derivation, a code, a plot, or a draft. Loops of this type power a growing body of agentic AI-for-science systems, e.g., \cite{boiko2023autonomous,lu2024aiscientist,ghafarollahi2024sciagents,gpd}. 

\Q{What can this buy us in July 2026?}
Concretely, use cases already include: literature reconnaissance with citations you then verify; reproduction of a paper's numerical baseline in an afternoon; generation of the ML boilerplate for every technique in this lecture, while you spend your time on the part that requires more attention from a physicist: the physics computation and how to verify it. The Day 2 results are a case study: several of those papers were produced with AI-assisted research workflows in the loop, with humans owning correctness. My research group's honest current estimate is that the implementation barrier for ML-for-strings projects has dropped by an order of magnitude, and  students who internalize this will be extremely productive.

Of course, there are \textbf{significant warnings}, which we must discuss:
\begin{itemize}
\item \textbf{Plausible $\neq$ correct.} A convincing wrong derivation is worse than no derivation. Treat agent output the way you treat a collaborator's: enthusiastically, but still with a critical eye. Check every step, at the source, whether code or calculation.
\item \textbf{Verify Results.} Demand unit tests, limiting cases, dimensional analysis, closed-form cross-checks (e.g., the DOZZ comparison of Day 2). An agent that writes code can also write the tests for it, but check it: reward hacking is real, e.g., \cite{amodei2016concrete,skalse2022reward}.
\item \textbf{Hallucinated scholarship.} Citations, attributions, and ``well-known results'' must be checked, always. There are good pipelines: for instance, require BibTeX pulled directly from INSPIRE, with clickable URLs.
\item \textbf{You are the physicist.} Agents lower the cost of \emph{doing} physics. They do not always have good taste, and they never bear responsibility for claimed results. 
\end{itemize}
These and other caveats are important to keep in mind as agentic techniques continue to develop in the next few years.

\medskip
A concrete instantiation of this pipeline, and the one my colleagues and I build and use, is \textbf{Get Physics Done} (GPD) \cite{gpd}, an open-source agentic AI physicist from Physical Superintelligence. GPD wraps the loop of Fig.~\ref{fig:agent_loop} into a \emph{physics} research pipeline: it scopes a problem and its assumptions, lays out a phased plan, dispatches specialist subagents to carry out derivations and numerical work, and runs a verification stage that checks results against the constraints nature imposes (e.g., dimensions, limiting cases, symmetries), while holding notation consistent across project phases. It bakes in the discipline urged above, including the INSPIRE-sourced citation checks used for this lecture. It can install as commands inside existing coding runtimes such as Claude Code or Codex, but also is in beta-testing as a standalone app.

With that framing, agents as force multipliers wrapped around classic techniques, let us learn some of the classic techniques.

\subsection{PINNs and Calabi-Yau Metrics}
\label{sec:cymetrics}

\subsubsection{Physics-Informed Neural Networks}

Suppose you want to solve a PDE
\begin{equation}
\cD[u](x) = 0 \quad \text{on } \Omega, \qquad \cB[u](x) = 0 \quad \text{on } \partial\Omega,
\end{equation}
with $\cD$ a (generally non-linear) differential operator and $\cB$ encoding boundary or other conditions. The \textbf{physics-informed neural network} (PINN) strategy \cite{raissi2019physics,karniadakis2021physics} is disarmingly simple: 
\CT{PINN}{Let a network $u_\theta$ model the solution}
regressing on $\theta$ to achieve the desired physics. 
This may be done by a \textbf{concrete algorithm:}
\begin{enumerate}
\item \textbf{Architecture.} Choose a network architecture $u_\theta$ with sufficient expressivity to represent the solution. 
\item \textbf{Sample.} Take points $\{x_i\} \subset \Omega$ and $\{x_j\}\subset\partial\Omega$.
\item \textbf{Regress.} Taking the loss to be 
\begin{equation}
\cL[\theta] = \frac{1}{n}\sum_{i=1}^{n} \big|\cD[u_\theta](x_i)\big|^2 + \lambda_{\cB}\,\frac{1}{m}\sum_{j=1}^m \big|\cB[u_\theta](x_j)\big|^2 + \cdots,
\end{equation} 
optimize parameters to minimize the loss and (approximately) solve the equations.
\end{enumerate}
Here the dots can include normalization constraints or any other differentiable demand you wish to make of the solution. The derivatives inside $\cD[u_\theta]$ come for free from automatic differentiation: the same machinery that powers backpropagation. Strengths of PINNs include expressivity and easy incorporation of various constraints. Standard weaknesses are loss-term balancing and the silent failure mode where the optimizer happily minimizes the wrong local minimum. A partial answer to the second is \emph{verified numerics}: pairing a network's approximate solution with a rigorous proof that a genuine solution exists nearby, as in recent computer-assisted work on PDE \cite{platt2026nirenberg} and the discovery of unstable singularities of the Euler equations \cite{wang2025unstable}. A parallel line in condensed matter and quantum chemistry uses neural-network variational ans\"atze to find quantum many-body ground states \cite{Carleo:2016svm,pfau2020ab,Hermann:2020xqs,Hibat-Allah:2020egr}.

A final important concept for PINNs is a \emph{surrogate loss}: a tractable, differentiable stand-in for an objective that is difficult to optimize directly. We will see an explicit example with CY metrics.

\subsubsection{Calabi-Yau Metrics}

String theory hands us an excellent PINN problem. Consider your favorite string theory compactified on a Calabi-Yau threefold. An essential result is Yau's theorem \cite{yau1978ricci}:
\begin{theorem}[Yau]
A compact K\"ahler manifold $X$ with $c_1(X)=0$ has, in each K\"ahler class, a unique Ricci-flat K\"ahler metric.
\end{theorem}
\noindent This is a truly remarkable result. We would like to state it as a 
\CT{Miracle 1}{Topology $\rightarrow$ Geometry. Do a simple topological check, get guaranteed existence of a Ricci-Flat metric.}
In fact, the check is so easy to perform that there are numerous construction algorithms for Calabi-Yau threefolds, leading to an \textbf{exponentially large number of geometries.} The Kreuzer--Skarke classification alone gives $473{,}800{,}776$ reflexive four-polytopes \cite{Kreuzer:2000xy}, whose Calabi--Yau hypersurfaces are bounded above by $\sim\!10^{428}$ \cite{Demirtas:2020dbm} and estimated to fall into $\sim\!10^{400}$ diffeomorphism classes \cite{Chandra:2023afu}; on the F-theory side a construction algorithm reaches an ensemble of $\tfrac{4}{3}\times 2.96\times 10^{755}$ geometries \cite{Halverson:2017ffz,Taylor:2017yqr}, generically carrying large axion populations \cite{Demirtas:2018akl,Fallon:2025lvn}. Despite this, we are faced with a 
\CT{Disturbing Fact}{We do not know \emph{any} Calabi-Yau metrics for compact threefolds with full $SU(3)$ holonomy.}
Yes, of course we know a CY metric on $T^6$, but not in the full $SU(3)$ holonomy case, a case of interest in $\mathcal{N}=1$  compactifications.

\medskip
It is remarkable we have been able to do so much string geometry in the Calabi-Yau setting without knowing the \emph{fundamental object on the space}! How have we gotten so far? Before proceeding to our ML questions and techniques, we would like to address this. Yes, we have existence, which is important. Yes, we can compute lots of topological quanta (intersection numbers, Hodge numbers, vector bundle cohomology and the like) that encode \emph{discrete} physical quantities such as charges, gauge groups, and particle generations. However, though these happen in string geometries, the computations are really computations in string \emph{topology}.

Of course, we do also compute \emph{continuous} quantities in CY compactifications, such as holomorphic gauge couplings, BPS particle masses, BPS brane tensions, etc. These are decidedly geometric computations, e.g. related to volumes of the CY and its submanifolds. But this just raises another question:
\Q{How can we compute volumes without the metric?}
It is due to seminal work of Harvey and Lawson \textcolor{ForestGreen}{\cite{Harvey:1982xk}} 
\CT{Miracle 2}{Calibrated geometry. We compute volumes of calibrated submanifolds by integrating calibration forms, e.g. the volume of a holomorphic curve $C$ is  
\begin{equation}
    \text{Vol(C)} = \int_C J,
\end{equation}
where $J$ is the K\"ahler form.}
Such calibrated submanifolds are volume minimizing in their homology class. Why can we compute volumes without the metric? The math answer is calibration, but the associated \textbf{physics reason is BPS-ness}. BPS objects are protected by supersymmetry, and are obtained by wrapping branes on calibrated submanifolds. Divisors, special Lagrangians, and holomorphic curves in CY3 are calibrated four-cycles, three-cycles, and two-cycles respectively, with associated calibration forms $J\wedge J$, $\text{Re}(\Omega)$, and $J$. 

As a counter-point that makes the connection between BPS-ness and calibrated geometry clear: though $G_2$ holonomy manifolds have calibrated three-cycles and four-cycles, they do not have calibrated two-cycles. In $G_2$ compactifications of M-theory, which are $\mathcal{N}=1$, the calibrated three-cycles and four-cycles give rise to BPS domain walls and strings by wrapping M5-branes, but there are no BPS particles from wrapped M2-branes. The absence of calibrated two-cycles in $G_2$ holonomy manifolds is the geometric reason for this. See \textcolor{ForestGreen}{\cite{Halverson:2015vta}} for an in-depth discussion of the matter in the $G_2$ case.

\bigskip
Since we do not have an \emph{expression} for the CY metric, we ask
\Q{Can a neural network \emph{approximate} a CY metric?}
 This matters physically, as we have just reviewed.
   There is a significant literature on numerical CY metrics: Donaldson's balanced-metric algorithm over algebraic ans\"atze \cite{Donaldson:2005hvr,Douglas:2006rr}, lattice and energy-functional methods of Headrick-Wiseman and Headrick-Nassar \cite{Headrick:2005ch,Headrick:2009jz}, and Donaldson-meets-ML hybrids \cite{Ashmore:2019wzb}. In 2020, neural networks took over as a new ansatz \cite{Anderson:2020hux,Douglas:2020hpv,Jejjala:2020wcc}, the current state-of-the-art. What topology and calibrated geometry alone cannot give us are the genuinely metric-dependent quantities: e.g., physical (canonically normalized) Yukawa couplings, matter-field K\"ahler metrics, and Kaluza-Klein modes. These are exactly what a numerical Calabi--Yau metric unlocks: once a network approximates the Ricci-flat metric one can compute normalized Yukawa couplings \cite{Anderson:2020hux,Butbaia:2024tje,Butbaia:2024xgj,Mishra:2025xkr}, bundle slopes and approximate Hermitian--Yang--Mills connections \cite{Larfors:2022nep,Ashmore:2019wzb}, Kaluza-Klein modes \cite{Ashmore:2020ujw}, and very likely many more observables in the coming years.

\medskip
Before doing a more clever version of the ML that leads to stronger results, we would like to do the obvious thing for the sake of pedagogy. The most literal PINN here outputs the metric components directly
\CT{Idea}{NN is the metric, $g^\theta_{\mu\bar\nu}(x)$, optimize $\theta$ to achieve $R_{\mu\bar\nu}=0$.}
Driving $R_{\mu\bar\nu}\to 0$ requires the Ricci tensor of the unconstrained $g^\theta$, a second-order expression in the metric.
The naive loss is then mean-squared-error on the Ricci components,
\begin{equation}
\label{eqn:ricci_loss}
\cL_{\text{Ricci}}[\theta] = \frac{1}{n}\sum_{i=1}^{n}
\sum_{\mu,\bar\nu}\big| R_{\mu\bar\nu}(x_i)\big|^2,
\end{equation}
 Two problems, though. First, the Ricci tensor is two derivatives above the metric, so each evaluation back-propagates through a noisy second-derivative stack. Second, \eqref{eqn:ricci_loss} neither keeps $g_\theta$ K\"ahler nor pins its K\"ahler class: the zero-loss set is degenerate (including the K\"ahler moduli space and non-K\"ahler Ricci-flat deformations), and the class can drift.

We can improve upon this by using details of Yau's proof. In a \emph{fixed} K\"ahler class, Ricci-flatness is equivalent to the complex Monge-Amp\`ere equation
\begin{equation}
\label{eqn:MA}
J_\text{CY} \wedge J_\text{CY} \wedge J_\text{CY} = \kappa \, \Omega \wedge \bar \Omega
\end{equation}
an equation for differential forms that must match point-by-point on the Calabi-Yau. The K\"ahler structure also tames the derivatives: the metric now descends from a single \textbf{K\"ahler potential} $K$ via 
\begin{equation}
g_{\mu\bar\nu}=\partial_\mu\partial_{\bar\nu}K
\end{equation} (so $J=i\,\partial\bar\partial K$), and the Ricci form \begin{equation}R_{\mu\bar\nu}=-\partial_\mu\partial_{\bar\nu}\log\det g\end{equation} then sits two derivatives above the metric and \emph{four} above the potential (which is why \cite{Larfors:2022nep} disable the Ricci loss by default), whereas the Monge-Amp\`ere surrogate constrains $J_\text{CY}$ directly, two derivative-orders cheaper. Writing
\begin{equation}J_\text{CY} = J_{\text{ref}} + \partial\bar\partial\phi\end{equation} builds in K\"ahlerity and fixes the class for free. K\"ahlerity, because $\partial\bar\partial\phi$ is automatically $d$-closed, so $J$ stays a closed $(1,1)$ form; and the class, because $\partial\bar\partial\phi$ is globally exact (with $\phi$ a genuine function), so $[J]=[J_{\text{ref}}]$ by the $\partial\bar\partial$-lemma: it cannot drift. Recalling that the K\"ahler form and metric are the same data, $J_{\mu\bar\nu}=i\,g_{\mu\bar\nu}$, the metric reads straight off the corrected form,
\begin{equation}
g_{\mu\bar\nu} = (g_{\text{ref}})_{\mu\bar\nu} + \partial_\mu\partial_{\bar\nu}\phi .
\end{equation}
Given a $J_\text{ref}$, one can model the correction with a neural network $\phi_\theta$.

\medskip
The approach of Larfors, Lukas, Ruehle, and Schneider \cite{Larfors:2022nep}, in the package \texttt{cymetric} \cite{cymetric}, does exactly this for complete-intersection (CICY) and Kreuzer-Skarke Calabi-Yaus, anywhere in moduli space. Related open-source packages include \texttt{cymyc} \cite{Butbaia:2024xgj} and \texttt{CYJAX} \cite{Gerdes:2022nzr}.
The \textbf{algorithm} has three essential components: an ansatz, a data sampler, and a loss function.

\medskip
\noindent\textbf{Ansatz.} Learn a correction to a reference; the $\phi$-model predicts
\begin{equation}
J_{\text{pred}} = J_{\text{ref}} + \partial\bar\partial\, \phi_\theta,
\end{equation}
with $J_{\text{ref}}$ (e.g.\ pulled-back Fubini-Study) fixing the class and the scalar network $\phi_\theta$ optimized during learning.  They also allow matrix-valued ans\"atze that do not ensure K\"ahlerity but instead restore it via a loss term. 

\medskip
\noindent\textbf{Data.} Sample points from $X$ by intersecting with random lines, where Shiffman-Zelditch-type theorems give the measure exactly. Integrals become weighted Monte Carlo sums,
\begin{equation}
\int_X \mathrm{dVol}_{CY}\, f \;\simeq\; \frac{1}{N}\sum_i \tilde w_i\,\det g(p_i)\, f(p_i),
\end{equation}
where the $p_i$ are the sampled points and Shiffman-Zelditch fixes their Fubini-Study sampling measure $\mathrm{d}A$ \emph{exactly}, so the weights $\tilde w_i$
are the known factors transferring that measure to the CY measure \cite{Larfors:2022nep}.

\medskip
\noindent\textbf{Losses.} Physics enters as the \texttt{cymetric} loss of \cite{Larfors:2022nep}, a weighted sum of residuals $\cL = \sum_k \alpha_k \cL_k$:
\begin{align}
\cL_{\text{MA}} &= \Big\lVert 1 - \tfrac{1}{\kappa}\tfrac{J_\text{CY}\wedge J_\text{CY} \wedge J_\text{CY}}{\Omega\wedge\bar\Omega}\Big\rVert, \label{eqn:cy_ma}\\
\cL_{\text{dJ}} &= \textstyle\sum_{ijk}\!\big(\lVert\mathrm{Re}\,c_{ijk}\rVert + \lVert\mathrm{Im}\,c_{ijk}\rVert\big),\ \ c_{ijk}=g_{i\bar j,k}-g_{k\bar j,i},\\
\cL_{\text{trans}} &= \tfrac1d\textstyle\sum_{\text{patches}}\big\lVert g^{(\cV)} - T\,g^{(\mathcal{U})}\,T^\dagger\big\rVert,\\
\cL_{\text{Kclass}} &= \tfrac{1}{h^{1,1}}\textstyle\sum_\alpha \big\lVert \mu(\cO(e_\alpha)) - \!\int_X\! J^2\wedge F_{\text{FS},\alpha}\big\rVert.
\end{align}
The Monge-Amp\`ere residual \eqref{eqn:cy_ma} enforces \eqref{eqn:MA}; $\cL_{\text{dJ}}$ and $\cL_{\text{trans}}$ restore K\"ahlerity and patch-consistency for ans\"atze that lack them; and the slope loss $\cL_{\text{Kclass}}$ is the global handle, pinning the K\"ahler class by matching computed line-bundle slopes $\mu(L)=\int_X J^2\wedge c_1(L)$ to their exact topological values. Trained this way, \cite{Larfors:2022nep} obtains metrics on the quintic, the bicubic, and a Kreuzer-Skarke manifold.

\subsubsection{Relation to Ricci Flow}

Training a neural network to be a metric is gradient descent in $\theta$-space, but it induces a \emph{flow in the space of metrics} as the network learns. A natural question:
\Q{Is Ricci flow a neural network metric flow?}
After all, the neural network techniques we've discussed and Ricci flow both have Ricci-flat fixed points.

\begin{figure}[ht]
\centering
\resizebox{0.92\columnwidth}{!}{%
\begin{tikzpicture}[every node/.style={font=\small}]
  \draw[rounded corners, thick] (0,0) rectangle (11,6.6);
  \node[anchor=north] at (5.5,6.45) {\textbf{NN metric flows} \,\footnotesize(non-local, stochastic, time-dependent)};
  \draw[rounded corners, thick] (0.7,0.5) rectangle (10.3,5.5);
  \node[anchor=north] at (5.5,5.35) {$\infty$-NN metric flows \,\footnotesize(frozen kernel $\bar\Theta$)};
  \draw[rounded corners, thick] (1.7,1.1) rectangle (9.3,4.4);
  \node[anchor=north] at (5.5,4.25) {Local metric flows \,\footnotesize($\bar\Theta\propto\delta$)};
  \draw[rounded corners, thick, fill=black!8] (2.9,1.8) rectangle (8.1,3.4);
  \node[anchor=north, align=center] at (5.5,3.25) {\textbf{Perelman's Ricci flow}\\ \footnotesize(Perelman-functional loss)};
\end{tikzpicture}}
\caption{Hierarchy of neural-network metric flows \cite{Halverson:2023ndu}. The general flow is non-local, stochastic, and time-dependent; infinite width freezes the kernel, locality collapses it to a delta function, and choosing Perelman's functional as the loss lands on Ricci flow.}
\label{fig:metric_flows}
\end{figure}
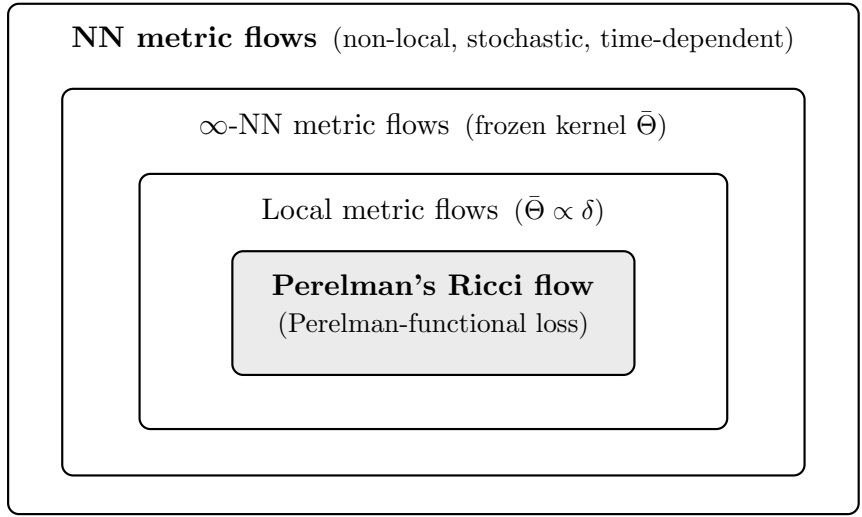

Day 1 gives the technology to answer the question \cite{Halverson:2023ndu}. The chain rule turns parameter-space gradient descent into a flow of the metric, the \textbf{metric flow}
\begin{equation}
\frac{d g_{ij}(x)}{dt} = -\int_X d\mu(x')\; \Theta_{ijkl}(x,x')\,\frac{\delta \ell(x')}{\delta g_{kl}(x')},
\end{equation}
driven by the functional derivative of the pointwise loss $\ell$ and governed by the \textbf{metric neural tangent kernel}
\begin{equation}
\Theta_{ijkl}(x,x') = \partial_{\theta_I} g_{ij}(x)\,\partial_{\theta_I} g_{kl}(x').
\end{equation}
In general this kernel is parameter-dependent, time-dependent, stochastic, and non-local, so the generic metric flow is \emph{not} Ricci flow. Ricci flow is a special corner, reached by three restrictions that parallel Day 1 exactly:
\begin{itemize}
\item \textbf{Infinite width.} As $N\to\infty$ the kernel freezes to a deterministic, time-independent $\bar\Theta$: no feature learning.
\item \textbf{Locality.} If the frozen kernel collapses to a delta function, $\bar\Theta_{ijkl}(x,x') = \delta(x-x')\,\delta_{ik}\delta_{jl}\,\bar\Omega(x)$, non-locality and index-mixing vanish and the flow becomes local, $d g_{ij}(x)/dt = -\bar\Omega(x)\,\delta\ell(x)/\delta g_{ij}(x)$.
\item \textbf{Perelman loss.} Choosing the loss to be (minus) Perelman's functional, $\ell = -\mathcal{F}[\phi,g]/\bar\Omega$ with $\mathcal{F}[\phi,g] = \int_X (R + |\nabla\phi|^2)\,e^{-\phi}\,dV$, the flow becomes Perelman's formulation of Ricci flow,
the very flow behind the proof of the $3d$ Poincar\'e conjecture.
\end{itemize}
So in general we have 
\CT{Answer}{Ricci flow $=$ special infinite-width NN metric flow.}
This is exemplified in  (Fig.~\ref{fig:metric_flows}) and explains why, in contrast, CY metric flows have been successful: they utilize feature learning.

\subsection{Reinforcement Learning and Search}
\label{sec:RL}

\subsubsection{RL Essentials}

Many problems in string theory exhibit significant structure and a vast configuration space. We will consider an abstract landscape.
\Q{How do you learn to search a vast landscape?}
It must happen according to what the landscape offers: generally an enormous configuration space with discrete and / or continuous quantities defining its states, a notion of ``good'' and ``bad'', and the ability to explore. That is the natural habitat of \textbf{reinforcement learning} \cite{sutton2018reinforcement} (RL).

A crisp notion in computer science formalizes the habitat:
\begin{definition}[Markov decision process (MDP)]
States $s\in\cS$, actions $a\in\cA$, a discount $\gamma\in[0,1)$, and dynamics given by a Markov transition kernel $P(s'|s,a)$ and reward $r(s,a)$. An MDP is the tuple $(\cS,\cA,P,r,\gamma)$.
\end{definition}
\noindent Reinforcement learning poses a control problem on top of the MDP: a \textbf{policy} $\pi_\theta(a|s)$ (for us, a neural network) is a state-conditioned distribution over actions (possibly deterministic). One would like to have a high return
\begin{equation}
J(\theta) = \bE_{\tau\sim\pi_\theta}\Big[\sum_{t\geq0} \gamma^t\, r(s_t,a_t)\Big], \qquad s_0\sim\mu_0,
\end{equation}
where the expectation is over trajectories $\tau=(s_0,a_0,s_1,a_1,\ldots)$ generated by the policy $\pi_\theta$ and the MDP dynamics, with $s_0$ drawn from some initial distribution $\mu_0$. The goal of RL is to find a policy that maximizes $J(\theta)$.
The value functions record expected return from a state, or a state-action pair:
\begin{equation}
\begin{aligned}
V^{\pi_\theta}(s) &= \bE_{\pi_\theta}\Big[\sum_{t\geq0}\gamma^t\, r(s_t,a_t) \;\Big|\; s_0=s\Big],\\
Q^{\pi_\theta}(s,a) &= \bE_{\pi_\theta}\Big[\sum_{t\geq0}\gamma^t\, r(s_t,a_t) \;\Big|\; s_0=s,\, a_0=a\Big],
\end{aligned}
\end{equation}
and the advantage 
\begin{equation}
A^{\pi_\theta}(s,a)=Q^{\pi_\theta}(s,a)-V^{\pi_\theta}(s)
\end{equation} measures how much better an action is than the policy's average: reinforce what worked, suppress what did not. 

The policy gradient theorem makes the objective trainable:
\begin{equation}
\nabla_\theta J(\theta) = \bE_{\tau\sim \pi_\theta}\Big[\sum_t \gamma^t\, \nabla_\theta \log \pi_\theta(a_t|s_t)\; Q^{\pi_\theta}(s_t,a_t)\Big].
\label{eqn:pg}
\end{equation}
$Q^{\pi_\theta}$ is often replaced by the advantage $A^{\pi_\theta}$ (equivalently, subtracting a state-dependent baseline), which leaves the expected gradient unchanged (subtracting any action-independent baseline $b(s_t)$ never changes it) and typically reduces variance. This is also what lets RL train on a non-differentiable reward: a black-box scorer or a discrete correctness check. One never differentiates the reward: by the \emph{log-derivative trick},
\begin{align}
\nabla_\theta\,\bE_{\tau\sim p_\theta}[R(\tau)] &= \bE_{\tau\sim p_\theta}\big[R(\tau)\,\nabla_\theta\log p_\theta(\tau)\big] \nonumber\\
&= \bE_{\tau\sim p_\theta}\Big[R(\tau)\sum_t \nabla_\theta\log\pi_\theta(a_t|s_t)\Big],
\end{align}
the entire gradient is carried by the differentiable log-probability $\log\pi_\theta$, with the return $R(\tau)$ (or $Q^{\pi_\theta}$, or the advantage) entering only as a scalar weight. Causality (rewards banked before time $t$ cannot depend on $a_t$) lets one replace $R(\tau)$ in the $t$-th term by the discounted reward-to-go from $t$, whose conditional expectation is $\gamma^t\, Q^{\pi_\theta}(s_t,a_t)$, recovering \eqref{eqn:pg}.

Three basic method families differ in what they parameterize:
\begin{itemize}
\item \textbf{Value-based.} Learn an action-value function, typically an approximation to the optimal value $Q^*$; the policy is then implicit: act \emph{greedily}, taking the highest-value action $\arg\max_a Q(s,a)$, or \emph{$\epsilon$-greedily}, taking that greedy action with probability $1-\epsilon$ and a uniformly random action with probability $\epsilon$ to keep exploring. In deep $Q$-learning one regresses $Q_\phi(s,a)$, by ordinary backprop through a temporal-difference loss, toward a bootstrapped target $r+\gamma\max_{a'}Q_{\bar\phi}(s',a')$, with the next-state estimate (from a target network $\bar\phi$) held fixed. E.g.\ $Q$-learning / deep $Q$-networks \cite{mnih2015human}.
\item \textbf{Policy-based.} Parameterize $\pi_\theta$ and optimize the policy objective \eqref{eqn:pg} directly. In the purest case, as in REINFORCE, no learned value function is required; in practice, methods such as TRPO often use baselines or advantage estimates while still treating the policy as the primary learned object. Gradients flow through $\log\pi_\theta$ by the log-derivative trick above. E.g.\ REINFORCE; trust-region methods such as TRPO \cite{schulman2015trust} that cap the per-step KL change.
\item \textbf{Actor-critic.} Learn both: the critic estimates $V^{\pi_\theta}$, $Q^{\pi_\theta}$, or the advantage by temporal-difference regression, and the actor updates $\pi_\theta$ by the policy gradient with the critic's estimate as the weight. The critic cuts the gradient's variance, at the cost of bias when it is inaccurate. E.g.\ A3C \cite{mnih2016asynchronous}, many environment copies updated asynchronously.
\end{itemize}
Deep RL's headline results (Atari from pixels \cite{mnih2015human}, Go and chess by self-play \cite{AlphaZero,silver2017mastering}) come from these families plus search, and the reasoning models of Sec.~\ref{sec:agents} are their newest descendants: RL on chains of thought, where the action is the next token and the reward scores the completed answer. The post-training workhorse here is \textbf{group-relative policy optimization} (GRPO) \cite{deepseek2025r1}. For each prompt it samples a \emph{group} of completions from the old rollout policy and scores each with the reward. The group supplies its own baseline: an output's advantage is its reward standardized within that prompt's group (mean subtracted, divided by the group standard deviation), so completions that beat the group average are reinforced and those below it suppressed, with no learned value function to estimate. The completion-level advantage then weights the token log-probability updates for that completion. The policy is updated with a PPO-style objective, with a KL penalty to a reference model.

We close with a key 
\Q{When does RL beat blind search?}
When the problem has \emph{learnable structure}: reward shaping that gives gradients toward consistency, and regularities in the solution set that a policy can internalize. Both phenomena appear, vividly, in the string theory applications.

\subsubsection{Exploring String Vacua}

In \cite{Halverson:2019tkf} we set deep RL loose on the landscape of type IIA compactifications with intersecting D6-branes on a toroidal orientifold. The environment: a state is a brane configuration; actions add or modify branes; rewards score progress toward string consistency (a system of coupled non-linear Diophantine equations) and toward Standard Model-like gauge sectors. Concretely, a stack $a$ of $N_a$ D6-branes wraps a special-Lagrangian three-cycle $\Pi_a\subset X$ (orientifold image $\Pi_a'$), and the agent must steer toward configurations satisfying the orientifold consistency conditions \cite{Blumenhagen:2005mu}:
\begin{itemize}
\item \textbf{RR tadpole cancellation}, an equation \emph{in homology}: the net D6-brane charge must cancel the O6-plane charge,
\begin{equation}
\sum_a N_a\big([\Pi_a]+[\Pi_a']\big) = 4\,[\Pi_{O6}] \ \in\ H_3(X,\bZ).
\end{equation}
\item \textbf{K-theory ($\bZ_2$) constraints}: the torsion charges invisible to homology must also vanish, the absence of a global Witten anomaly, detected by $Sp$-probe branes.
\item \textbf{Supersymmetry}: each $\Pi_a$ must be a special Lagrangian three-cycle, i.e.\ calibrated by $\mathrm{Re}\,\Omega$:
\begin{equation}
J\big|_{\Pi_a}=0,\qquad \mathrm{Im}\,\Omega\big|_{\Pi_a}=0,
\end{equation}
so that $\mathrm{Re}\,\Omega$ restricts to the volume form on $\Pi_a$. A single such brane preserves $\cN=1$; for \emph{all} stacks (and the O6-plane) to preserve a \emph{common} $\cN=1$ they must be calibrated with respect to the same $\mathrm{Re}\,\Omega$, the same phase. On the factorized torus $T^6=\prod_{I=1}^3 T^2_I$ a stack wraps each $T^2_I$ at an angle $\theta_a^I$ to the O6-plane, its calibration phase is $\sum_I\theta_a^I$, and this common-phase condition reduces to the angle-sum condition
\begin{equation}
\sum_{I=1}^{3}\theta_a^I = 0 \pmod{2\pi},
\end{equation}
with $\theta_a^I$ the wrapping angle in the $I$-th $T^2$.
\item \textbf{Standard Model-like data} (the target, not a consistency requirement): chiral matter in the bifundamental $(N_a,\bar N_b)$ is counted by the topological intersection number $I_{ab}=[\Pi_a]\cdot[\Pi_b]$, so an SM-like spectrum (gauge group $\supseteq SU(3)\times SU(2)\times U(1)$ with three chiral generations) pins down the relevant $I_{ab}=\pm 3$.
\end{itemize}
A representative reward shape (the exact shaping in \cite{Halverson:2019tkf} differs, this is schematic) penalizes constraint violation and rewards hitting the target,
\begin{equation}
R \;=\; -\sum_a w_a\,\big|\,\Delta_a\,\big| \;+\; (\text{bonus when SM-like}),
\end{equation}
where the $\Delta_a$ are the (signed) violations of the constraints above and the $w_a$ are weights, so $R$ rises toward zero as the configuration approaches consistency. 

The IIA models explored with an A3C agent \cite{mnih2016asynchronous} that interacts with this environment, updating its policy and value networks from $R$, and learns. The outcomes worth remembering:
\begin{itemize}
\item \textbf{Efficiency.} After training, the agent finds consistent models at a rate up to $O(200)\times$ that of a random walker.
\item \textbf{Rediscovery of human strategy.} On one task the agent's learned policy reproduced a strategy familiar from human model-building (so-called filler branes saturating tadpoles): it found our tricks on its own.
\item \textbf{Genuinely new strategy.} On another task, with no known human-derived strategy, the agent invented one, achieving the goal twice as efficiently as the human-inspired baseline.
\end{itemize}
The agent is not doing magic; it is doing statistics on the solution set of Diophantine equations and caching the structure in its policy. But ``learned the structure of a corner of the landscape'' is exactly the capability one wants.
\CT{Key Point}{Given learnable structure, RL does not merely search faster --- it discovers new strategies.}
Related search techniques in the landscape include genetic algorithms \cite{Ruehle:2017mzq, Abel:2014xta}. A number of works \cite{Krippendorf:2021uxu,Cole:2021nnt,Abel:2021rrj,Abel:2021ddu,Constantin:2021for} utilize RL and / or GA, and sometimes test them against each other. See \cite{Ruehle:2020jrk} for a broader survey.

\subsubsection{Learning to Unknot and the Poincar\'e Conjecture}

RL is not only for landscapes: it can rigorously establish properties of objects in math and physics, in the certificate sense. We review a case in low-dimensional topology \cite{Gukov:2020qaj}. 

The UNKNOT problem asks whether a given knot is the unknot, and it can be made into a problem about \emph{language}. A knot is presented as a \textbf{braid word}: a string over the finite alphabet $\{\sigma_1^{\pm1},\dots,\sigma_{n-1}^{\pm1}\}$ of Artin generators. Deciding UNKNOT becomes \emph{sequence classification}, and the moves that simplify a braid (below) are a rewriting grammar on these strings. Posed as language, it yields to language models: Transformer architectures (Reformer, shared-QK) classify unknottedness from braid words with high accuracy (curiously \emph{increasing} with braid length) while learning, unprompted, an internal correlate of the Jones polynomial, their confidence tracking its degree. Hold that thought for Sec.~\ref{sec:interp}.

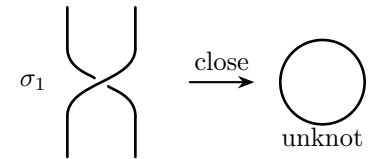
\begin{figure}[ht]
\centering
\begin{tikzpicture}[scale=0.9, line cap=round, line width=1pt]
  \draw (0,0) -- (0,0.6);
  \draw (1,0) -- (1,0.6);
  \draw (0,0.6) .. controls (0,1.1) and (1,1.1) .. (1,1.6);
  \draw (1,0.6) .. controls (1,0.9) and (0.66,0.99) .. (0.6,1.04);
  \draw (0.4,1.16) .. controls (0.34,1.21) and (0,1.3) .. (0,1.6);
  \draw (0,1.6) -- (0,2.2);
  \draw (1,1.6) -- (1,2.2);
  \node[font=\footnotesize] at (-0.5,1.1) {$\sigma_1$};
  \draw[-{Stealth[length=2.2mm]}, line width=0.8pt] (1.8,1.1) -- (2.75,1.1);
  \node[font=\footnotesize] at (2.27,1.42) {close};
  \draw (3.75,1.1) circle (0.62);
  \node[font=\footnotesize] at (3.75,0.3) {unknot};
\end{tikzpicture}
\caption{The Artin generator $\sigma_1 \in B_2$: two strands with one crossing. Closing the braid (joining each strand's top to its own bottom) gives a single loop, the unknot.}
\label{fig:unknot-braid}
\end{figure}

A braid on $n$ strands is a product of \emph{Artin generators} $\sigma_i$ (strand $i$ crosses over strand $i{+}1$) and their inverses $\sigma_i^{-1}$; the \emph{closure} (join each strand's top to its own bottom) turns a braid into a knot or link. The simplest example is the one-crossing two-strand braid $\sigma_1 \in B_2$ (Fig.~\ref{fig:unknot-braid}): its closure is the \textbf{unknot}, since the lone crossing undoes by a Reidemeister-I move. The catch is that much \emph{longer} words can also close to the unknot, in ways no short calculation reveals, which is exactly why deciding UNKNOT is hard, and why the RL ``simplify-by-moves'' framing below is a good  one.

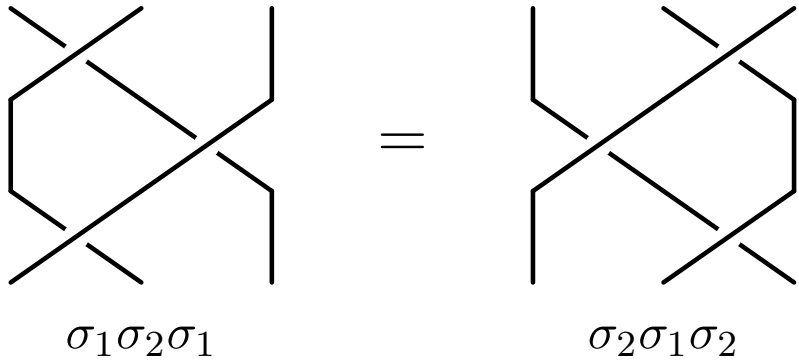
\begin{figure}[ht]
\centering
\resizebox{0.85\columnwidth}{!}{%
\begin{tikzpicture}[line cap=round, line width=1pt]
  \draw (1,0) -- (0,0.7); \fill[white] (0.5,0.35) circle (0.1); \draw (0,0) -- (1,0.7); \draw (2,0) -- (2,0.7);
  \draw (2,0.7) -- (1,1.4); \fill[white] (1.5,1.05) circle (0.1); \draw (1,0.7) -- (2,1.4); \draw (0,0.7) -- (0,1.4);
  \draw (1,1.4) -- (0,2.1); \fill[white] (0.5,1.75) circle (0.1); \draw (0,1.4) -- (1,2.1); \draw (2,1.4) -- (2,2.1);
  \node[font=\small] at (1,-0.45) {$\sigma_1\sigma_2\sigma_1$};
  \node[font=\large] at (3,1.05) {$=$};
  \draw (6,0) -- (5,0.7); \fill[white] (5.5,0.35) circle (0.1); \draw (5,0) -- (6,0.7); \draw (4,0) -- (4,0.7);
  \draw (5,0.7) -- (4,1.4); \fill[white] (4.5,1.05) circle (0.1); \draw (4,0.7) -- (5,1.4); \draw (6,0.7) -- (6,1.4);
  \draw (6,1.4) -- (5,2.1); \fill[white] (5.5,1.75) circle (0.1); \draw (5,1.4) -- (6,2.1); \draw (4,1.4) -- (4,2.1);
  \node[font=\small] at (5,-0.45) {$\sigma_2\sigma_1\sigma_2$};
\end{tikzpicture}}
\caption{The braid relation $\sigma_i\sigma_{i+1}\sigma_i=\sigma_{i+1}\sigma_i\sigma_{i+1}$ (here $i{=}1$ on three strands), the Reidemeister-III move in braid-word form, and the move the trained agent relies on most. The remaining move types act on the word the same algebraic way: far commutation $\sigma_i\sigma_j=\sigma_j\sigma_i$, free cancellation $\sigma_i\sigma_i^{-1}=1$ (Reidemeister II), and the Markov (de)stabilization (Reidemeister I).}
\label{fig:braid-moves}
\end{figure}

Let me explain the RL result. Two families of moves connect braid words with the same closure (Fig.~\ref{fig:braid-moves}): the \textbf{braid relations} within $B_n$, and the \textbf{Markov moves} relating different $B_n$,
\begin{equation}
\begin{aligned}
&\text{Braid:} && \sigma_i\sigma_{i+1}\sigma_i=\sigma_{i+1}\sigma_i\sigma_{i+1},\ \ \sigma_i\sigma_j=\sigma_j\sigma_i\ (|i{-}j|\geq2),\\
&\text{Free:} && \sigma_i\sigma_i^{-1}=1,\\
&\text{Markov:} && w\simeq g\,w\,g^{-1},\ \ w\simeq w\,\sigma_n^{\pm1}\ \text{((de)stabilization)},
\end{aligned}
\end{equation}
where destabilization, free cancellation, and the braid relation are the braid-word avatars of Reidemeister moves I, II, and III. An unknot is certified by a sequence of these reducing the word to the empty one. Cast move-selection as an MDP and train with TRPO \cite{schulman2015trust}: the agent learns to \emph{simplify}, producing explicit unknotting sequences (human-checkable certificates, not mere classifications) consistently across a wide range of crossing numbers where other RL algorithms and random walkers fail. Interrogating the learned policy is itself informative: the agent leans on braid relations more than on one of the Markov moves, a statement about the geometry of unknotting that emerged from training statistics. This certificate-producing pattern --- search hard, verify trivially --- is the template for ML results that pure mathematicians can accept at full rigor \cite{gukov2024rigor}.

The same approach can be brought to bear on the smooth four-dimensional Poincar\'e conjecture (SPC4), a major open problem in topology. A known strategy to disprove it involves pairs of knots with certain combinations of properties related to zero-surgeries and sliceness. Specifically, if $(K_1,K_2)$ are knots with the same 0-surgeries and exactly one is slice, then SPC4 is false. In \cite{Gukov_2025} the demonstration of sliceness is handed to Bayesian optimization and RL, and if the above property were established it would constitute a rigorous disproof of SPC4.  Applied to a family of $3375$ candidates \cite{MP21}, the programs eliminated all but two from the list, ruling out many proposed counterexamples to SPC4, by showing that the pairs were both not-slice or both slice.

\subsection{Conjecture Generation and Rigor}
\label{sec:interp}

\subsubsection{From Prediction to Understanding}

Supervised learning (Day 1) gives you a function that predicts, but with error. We often want more:
\Q{Can we interpret ML to get theorems?}
The mode of use is \textbf{conjecture generation} \cite{Carifio:2017bov}: a trained network is a source of candidate patterns, which the scientist then tries to state precisely and prove.
The proposal that ML could drive the discovery loop this way was made early in the string-landscape context, with further string-theoretic examples (conjectured formulas for line bundle cohomology \cite{Brodie:2019dfx} and interpretable models for F-theory matter spectra \cite{Bies:2020gvf}) followed by a striking demonstration in pure mathematics by DeepMind \cite{davies2021advancing}.
The working pipeline:
\begin{center}
\resizebox{\columnwidth}{!}{%
\begin{tikzpicture}[
  stage/.style={draw, thick, rounded corners, align=center, minimum height=0.8cm, font=\small},
  arr/.style={-{Stealth[length=2.2mm]}, thick}
]
\node[stage] (data) at (0,0) {data};
\node[stage] (model) at (2.2,0) {model};
\node[stage] (interp) at (4.6,0) {interpret};
\node[stage] (conj) at (7.3,0) {conjecture};
\node[stage] (proof) at (10,0) {proof};
\draw[arr] (data) -- (model);
\draw[arr] (model) -- (interp);
\draw[arr] (interp) -- (conj);
\draw[arr] (conj) -- (proof);
\draw[arr, dashed] (conj) to[bend right=25] node[above, font=\footnotesize]{refine} (data);
\end{tikzpicture}}
\end{center}
The interpret stage has a toolbox according to details of the ML algorithm:
\begin{itemize}
\item \textbf{Interpretable model classes.} Linear and logistic regression, decision trees, and small forests expose weights, splits, and feature importances directly. Often these match deep networks on structured math data; always try the simple thing first.
\item \textbf{Attribution.} For deep networks, gradient-based saliency ranks input features by influence: which invariants is the network actually using?
\item \textbf{Interpretable-by-design architectures.} KANs (Day 1) learn univariate activations that can be read off and snapped to symbolic form \cite{KAN}, and symbolic regression distills trained networks into formulas.
\item \textbf{Programs as hypotheses.} The LLM-era variant: search over \emph{code} rather than weights \cite{romera2024funsearch,novikov2025alphaevolve}, so the discovered object is interpretable by construction: you read the program.
\end{itemize}

\subsubsection{Case Studies in Conjecture Generation}

\noindent\emph{Gauge groups in the F-theory ensemble.} In \cite{Carifio:2017bov} we ran conjecture generation over an ensemble of
\begin{equation}
N \;=\; \tfrac{4}{3}\times 2.96\times 10^{755}
\end{equation}
F-theory compactifications, keeping models simple enough to interpret. A \textbf{decision tree} predicts the number of weak Fano toric threefolds from reflexive polytopes (each gives a smooth F-theory geometry); \textbf{linear regression} re-derived a known conjecture for the gauge-group \emph{rank}; and \textbf{logistic regression} produced a \emph{new} conjecture for \emph{when} an $E_6$ sector arises, which we then \emph{proved} \cite{gukov2024rigor}.
\medskip
\noindent\emph{Knot invariants at DeepMind.} The same loop, run by mathematicians \cite{davies2021advancing}: train a network to predict a hyperbolic knot's signature $\sigma(K)$ from its geometric invariants, use gradient saliency to find that only a handful matter, and distill them into a single new invariant, the \emph{natural slope} $\mathrm{slope}(K)$ read off from the cusp geometry. The network had effectively conjectured $\mathrm{slope}(K)\approx 2\,\sigma(K)$, made rigorous \cite{davies2021signature} as
\begin{equation}
\big|\,2\,\sigma(K) - \mathrm{slope}(K)\,\big| \;\leq\; c\,\frac{\mathrm{vol}(K)}{\mathrm{inj}(K)^3},
\end{equation}
with $\mathrm{vol}(K)$ the hyperbolic volume, $\mathrm{inj}(K)$ the injectivity radius, and $c$ universal: a new theorem in knot theory, with the network's role being to point the mathematicians at the right invariants.

\medskip
\noindent\emph{The Jones polynomial, rediscovered.} Recall from Sec.~\ref{sec:RL} that unknot-classifying networks spontaneously correlated their confidence with the degree of the Jones polynomial \cite{Gukov:2020qaj}: a known invariant, rediscovered inside the weights as the natural coordinate for the task. Interpretability sometimes tells you the network found something you know (reassuring), and sometimes that it found something you do not (a lead).

\medskip
\noindent\emph{Programs and agents.} FunSearch \cite{romera2024funsearch} found record cap-set constructions by evolving \emph{programs} scored by an exact verifier; AlphaEvolve \cite{novikov2025alphaevolve} broke a longstanding matrix-multiplication record the same way. Generator proposes interpretable programs, trusted verifier scores them, evolution iterates. This is conjecture generation driven by the agent loop of Sec.~\ref{sec:agents}, applicable wherever string theory offers a fast exact check: consistency conditions, index computations, topological invariants. A natural project for you: pick a landscape structure that was painstakingly discovered by hand, and ask whether a program-search agent finds it, or its generalization.

\section{Recap and Outlook}
\label{sec:recap}

Let me review the results of the three days and provide an outlook.

\textbf{Day 1} built neural network theory on three pillars. The first was expressivity: approximation theorems such as the UAT and the Kolmogorov-Arnold representation theorem show that networks can represent essentially anything, and they still inspire architectures today, as in the KANs. Attention and the transformer extend the story from fixed inputs to sequences, and chain-of-thought reasoning shows that expressive power grows with the number of tokens a model is allowed to emit, which is the engine beneath the reasoning models. The second pillar was statistics: a freshly initialized network is a single draw from an ensemble, and the correlators of that ensemble are its fundamental objects. At large $N$ the central limit theorem renders the ensemble a generalized free field (the NNGP), interactions are restored either at $O(1/N)$ or by breaking statistical independence, and global symmetries can be engineered by absorbing transformations into the parameter density. The third pillar was dynamics: gradient descent is governed by the neural tangent kernel, which at infinite width freezes into exactly solvable kernel regression. The glaring caveat is that the frozen kernel has no feature learning, and the $\mu P$ scaling analysis shows how to restore it.

\textbf{Day 2} turned the correspondence around and defined a field theory by an architecture and a parameter density, $(\phi_\theta, P(\theta))$, with the partition function an integral over parameters rather than fields. We saw how to engineer free theories via the CLT and spectrum shaping, how to introduce interactions via operator insertions that deform $P(\theta)$ and break independence, and how quantumness becomes a checklist of OS axioms that one engineers towards. The second half covered results from the last year. Universality states that \emph{every} distribution over tempered distributions is an NN-FT. Liouville theory was realized with all of the interactions carried by a conditioned zero mode, matching DOZZ to the percent level. The bosonic string was realized at infinite width, with $\alpha'$ set by the output-weight variance and the Veneziano and Virasoro-Shapiro amplitudes recomputed analytically in parameter space. Topological sectors entered through discrete latents, yielding vortices and the BKT transition. Finally, Schwinger-Dyson equations, Ward identities, and anomalies were formulated as flows on parameter space, including a mode-counting NN-FT derivation of $D=26$, the critical dimension of the bosonic string.

\textbf{Day 3} was the applied ML-for-physics day. Agents, reasoning models acting through tool loops, have significantly lowered the implementation cost of every classic technique, provided their outputs are carefully verified. PINNs solve geometric PDEs by descending residuals, with Calabi-Yau metrics as the central example in string theory; the NTK of Day 1 reappeared there, connecting network metric flows to Ricci flow in the frozen-kernel limit. Reinforcement learning turns landscape exploration and unknotting into learned search, and it produces not only efficiency but also human-checkable certificates; the applications were in type II string theory and knot theory. Finally, interpretable supervised learning can lead to theorems via conjecture generation and subsequent refinement, as we saw for F-theory gauge sectors and knot signatures, with program-search agents as a modern variant.

\bigskip

In my opinion, our agentic era sets a course for the coming years. The techniques of Day 3 were developed by researchers who paid the time cost of implementation, and the students in this school largely will not \emph{have} to, which changes what a single researcher can attempt. Over the next few years agents will take on more and more of the research pipeline, from literature synthesis through computation and first-pass verification, becoming more reliable as formal and numerical checks are built into their loops. I also expect conjecture generation to become increasingly agentic, with systems that propose, test, and refine candidate structures against the kind of exact, noiseless data that string theory has in abundance. Time saved on tedious calculation can be reinvested in deeper thinking and the pursuit of conceptual breakthroughs. That mode of work is why many of us became physicists in the first place, and it is a skill worth training deliberately. It is an exciting time to be a physicist.

\vspace{.5cm}
\noindent \textbf{Acknowledgements:} I would like to thank SIMIS and the organizers and students of the Pre-Strings 2026 school \cite{prestrings2026} for the invitation and for an energetic time in Shanghai. These lectures lean on the work of and conversations with many friends and collaborators, especially Mehmet Demirtas, Christian Ferko, Sam Frank, Vishnu Jejjala,  Anindita Maiti, Brandon Robinson, Fabian Ruehle,  Matt Schwartz, and Keegan Stoner on the NN-FT results of Day 2; Yasaman Bahri, Cengiz Pehlevan, and Greg Yang on the ML theory of Day 1; and many others on work appearing throughout. I would like to thank Christian Ferko, Sam Frank, and Fabian Ruehle for reading draft sections of these lectures. My works presented in these lectures are supported by the National Science Foundation under CAREER grant PHY-1848089, grant PHY-2209903, and Cooperative Agreement PHY-2019786 (The NSF AI Institute for Artificial Intelligence and Fundamental Interactions), as well as the Simons Collaboration on the Physics of Learning and Neural Computation. I thank my colleagues at Physical Superintelligence for discussions and collaboration on agentic AI-for-Physics. In the spirit of those methods and Day 3 lectures, I worked with Claude Fable 5 to draft these notes, and spent many hours afterward adding and polishing content; some of the content is pulled from my TASI lectures. Two-column landscape mode is chosen to facilitate boardwork, and because it remains a pleasant read, HT @ \cite{ginsparg1988applied}.

\vspace{.5cm}
\noindent \textbf{Disclaimers:} I am posting these notes shortly after the school itself. Therefore, they surely contain more typos and fewer references than is ideal; corrections are welcome and will be folded into updates. I struck a playful tone, because lectures should be fun, but may have overdone it in places.  The treatment of many results is necessarily compressed, and I refer the reader to the original papers.

\appendix
\section{Central Limit Theorem}
\label{app:CLT}
Let us recall a quick derivation of the Central Limit Theorem (CLT), since the statistics of wide neural networks hinges on it. Take a sum of random variables
\begin{equation}
\phi = \frac{1}{\sqrt{N}} \sum_{i=1}^N X_i,
\end{equation}
with $\bk{X_i} = 0$. Moments $\mu_r$ and cumulants $\kappa_r$ come from the moment generating function (partition function) $Z[J] = \bk{e^{J\phi}}$ and the cumulant generating function $W[J] = \log Z[J]$:
\begin{align}
    \mu_r & = \left(\frac{d}{dJ}\right)^r Z[J]\bigg|_{J=0} \\
    \kappa_r & = \left(\frac{d}{dJ}\right)^r W[J]\bigg|_{J=0}.
\end{align}
For independent $X_i$ the partition function factorizes,
$Z_{\sum_i X_i}[J] = \prod_i Z_{X_i}[J]$, so cumulant generating functions add:
\begin{align}
    W_{\sum_i X_i}[J] = \sum_i W_{X_i}[J] \label{eqn:Windep}\\
    \kappa_r^{\sum X_i} = \sum_i \kappa_r^{X_i}.
\end{align}
If the $X_i$ are also identically distributed, all the $\kappa_r^{X_i}$ coincide, and tracking the $1/\sqrt{N}$ normalization gives
\begin{equation}
\kappa_r^\phi = \frac{\kappa_r^{X_i}}{N^{r/2-1}},
\end{equation}
whence
\begin{equation}
\lim_{N\to\infty} \kappa_{r>2}^\phi = 0. \label{eqn:CLT_kappar}
\end{equation}
Vanishing higher cumulants is precisely Gaussianity. In physics language: cumulants are connected correlators, and \eqref{eqn:CLT_kappar} says free theories have no connected higher-point functions.

The derivation exposes exactly two ways to generate non-Gaussianity:
\begin{itemize}
    \item \textbf{$1/N$-corrections}, visible in $\kappa_r^\phi$ at finite $N$.
    \item \textbf{Independence breaking}, since the proof leaned on \eqref{eqn:Windep}.
\end{itemize}
Both will be put to work repeatedly in these lectures.

\bibliographystyle{JHEP}
\bibliography{refs_clean_verified_all}

\end{document}